\documentclass[12pt,a4paper,oneside,headsepline,final]{scrbook}
\usepackage{lmodern}
\usepackage{setspace}
\usepackage{amssymb,amsmath}
\usepackage{ifxetex,ifluatex}
\usepackage{fixltx2e} 
\ifnum 0\ifxetex 1\fi\ifluatex 1\fi=0 
  \usepackage[T1]{fontenc}
  \usepackage[utf8]{inputenc}
\else 
  \ifxetex
    \usepackage{mathspec}
  \else
    \usepackage{fontspec}
  \fi
  \defaultfontfeatures{Mapping=tex-text,Scale=MatchLowercase}
  
\fi
\IfFileExists{upquote.sty}{\usepackage{upquote}}{}
\IfFileExists{microtype.sty}{%
\usepackage{microtype}
\UseMicrotypeSet[protrusion]{basicmath} 
}{}
\usepackage[inner=4cm,a4paper]{geometry}
\makeatletter
\@ifpackageloaded{hyperref}{}{%
\ifxetex
  \usepackage[setpagesize=false, 
              unicode=false, 
              xetex]{hyperref}
\else
  \usepackage[unicode=true]{hyperref}
\fi
}
\@ifpackageloaded{color}{
    \PassOptionsToPackage{usenames,dvipsnames}{color}
}{%
    \usepackage[usenames,dvipsnames]{color}
}
\makeatother
\hypersetup{
            breaklinks=true,
            pdfauthor={Jonas Weber},
            pdftitle={Idiomatic and Reproducible Software Builds using Containers for Reliable
Computing},
            colorlinks=true,
            citecolor=black,
            urlcolor=black,
            linkcolor=black,
            pdfborder={0 0 0}
            }
\usepackage{listings}
\usepackage{longtable,booktabs}
\usepackage{graphicx,grffile}
\usepackage{xcolor}
\usepackage{tikz}
\usepackage{rotating}
\usepackage{multirow}
\usepackage{capt-of}
\usepackage{pgfplots}
\makeatletter
\def\maxwidth{\ifdim\Gin@nat@width>\linewidth\linewidth\else\Gin@nat@width\fi}
\def\maxheight{\ifdim\Gin@nat@height>\textheight\textheight\else\Gin@nat@height\fi}
\makeatother
\setkeys{Gin}{width=0.9\maxwidth,height=0.9\maxheight,keepaspectratio}
\providecommand{\tightlist}{%
  \setlength{\itemsep}{0pt}\setlength{\parskip}{0pt}}

\title{Idiomatic and Reproducible Software Builds using Containers for Reliable
Computing}
\author{Jonas Weber}
\date{April 18, 2016}

\usepackage{scrhack}

\usepackage{amsthm}
\theoremstyle{plain}

\theoremstyle{definition}
\newtheorem{definition}{Definition}[section]

\lstdefinelanguage{JavaScript}{
  keywords={break, case, catch, continue, debugger, default, delete, do, else, false, finally, for, function, if, in, instanceof, new, null, return, switch, this, throw, true, try, typeof, var, void, while, with},
    morecomment=[l]{//},
    morecomment=[s]{/*}{*/},
    morestring=[b]',
    morestring=[b]",
    ndkeywords={class, export, boolean, throw, implements, import, this},
    sensitive=true
}

\lstdefinelanguage{Dockerfile}{
  keywords={FROM,MAINTAINER,RUN,CMD,LABEL,EXPOSE,ENV,ADD,COPY,ENTRYPOINT,VOLUME,USER,WORKDIR,ONBUILD},
    morecomment=[l]{\#},
    moredelim=[s][\bfseries]{\\\$\{}{\}},
    morestring=[b]",
    showstringspaces=false,
    sensitive=false,
    numbers=right,
    numberstyle=\tiny
}

\lstdefinelanguage{Go}{
    morekeywords=[1]{break,default,func,interface,%
      case,defer,go,map,struct,chan,else,goto,package,%
        switch,const,fallthrough,if,range,type,continue,%
          for,import,return,var,select},
      morekeywords=[2]{make,new,nil,len,cap,copy,complex,%
        real,imag,panic,recover,print,println,iota,close,%
          closed,_,true,false,append,delete},
      morekeywords=[3]{%
        string,int,uint,uintptr,double,float,byte,%
          int8,int16,int32,int64,int128,%
          uint8,uint16,uint32,uint64,uint128,%
          float32,float64,complex64,complex128,%
          rune},
      morestring=[b]{"},
    morestring=[b]{'},
    morestring=[b]{`},
      comment=[l]{//},
    morecomment=[s]{/*}{*/},
      sensitive=true,
  numbers=right,
  numberstyle=\tiny
}

\lstdefinestyle{mylua}
{
  language         = {[5.2]Lua},
  showstringspaces = false,
  numbers=right,
  numberstyle=\tiny
}

\newcommand\YAMLcolonstyle{\mdseries}
\newcommand\YAMLkeystyle{\bfseries}
\newcommand\YAMLvaluestyle{\mdseries}

\makeatletter

\newcommand\language@yaml{yaml}

\expandafter\expandafter\expandafter\lstdefinelanguage
\expandafter{\language@yaml}
{
  keywords={true,false,null,y,n},
  keywordstyle=\bfseries,
  basicstyle=\ttfamily\YAMLkeystyle,                                 
  numbers=right,
  numberstyle=\tiny,
  sensitive=false,
  comment=[l]{\#},
  morecomment=[s]{/*}{*/},
  commentstyle=\ttfamily,
  stringstyle=\YAMLvaluestyle\ttfamily,
  moredelim=[l][\color{orange}]{\&},
  moredelim=[l][\color{magenta}]{*},
  moredelim=**[il][\YAMLcolonstyle{:}\YAMLvaluestyle]{:},   
  morestring=[b]',
  morestring=[b]",
  literate =    {---}{{\ProcessThreeDashes}}3
                {>}{{\textcolor{red}\textgreater}}1
                {|}{{\textcolor{red}\textbar}}1
                {\ -\ }{{\mdseries\ -\ }}3,
}

\lst@AddToHook{EveryLine}{\ifx\lst@language\language@yaml\YAMLkeystyle\fi}
\makeatother

\usepackage{floatrow}
\makeatletter
\def\maketitle{%
\begin{titlepage}
	\begin{center}
		\mbox{}

		\vspace{1.5cm}

    {\large\textsc{Master's Thesis}}

		\vspace{1.5cm}

		{\huge\bfseries \@title}

		\vspace{1.5cm}

		Jonas Weber

    \@date

    \vfill

		{\large Albert-Ludwigs-Universität Freiburg}\\
		{\large Faculty of Engineering}\\
		{\large Department of Computer Science}\\
		{\large Bioinformatics}\\
	\end{center}
  \newpage\thispagestyle{empty}

  Eingereichte Masterarbeit gemäß den Bestimmungen der Prüfungsordnung der Albert-Ludwidgs-Universität Freiburg
  für den Studiengang Master of Science (M.Sc.) Informatik vom 19. August 2005.

  \vfill

  \setstretch{1}

  \textbf{Bearbeitungszeitraum}\\
  12. Januar 2016 - 12. Juli 2016

	\bigskip

  \textbf{Gutachter}\\
  Prof. Dr. Rolf Backofen\\
  Head of the Group\\
  Chair for Bioinformatics\\

  \textbf{Zweitgutachter}\\
  Prof. Dr. Christoph Scholl\\
  Director\\
  Chair of Operating Systems\\

	\bigskip

  \textbf{Betreuer}\\
  Dr. Björn Grüning

  \setstretch{1.5}
\end{titlepage}
}
\makeatother

\usepackage[automark]{scrlayer-scrpage}
\automark{chapter}
\automark*{section}
\clearpairofpagestyles
\ihead{\headmark}
\ohead{\pagemark}

\usepackage[defernumbers,sorting=none]{biblatex}
\DeclareBibliographyCategory{cited}
\AtEveryCitekey{\addtocategory{cited}{\thefield{entrykey}}}

\begin{document}
\selectcolormodel{gray}
\maketitle

\hypersetup{linkcolor=black}
\setcounter{tocdepth}{2}

\addtocontents{toc}{\protect\setcounter{tocdepth}{-1}}

\chapter*{Abstract}\label{abstract}
\addcontentsline{toc}{chapter}{Abstract}

Containers as the unit of application delivery are the `next big thing'
in the software development world. They enable developers to create an
executable image containing an application bundled with all its
dependencies which a user can run inside a controlled environment with
virtualized resources. Complex workflows for business-critical
applications and research environments require a high degree of
reproducibility which can be accomplished using uniquely identified
images as units of computation.

It will be shown in this thesis that the most widely used approaches to
create an image from pre-existing software or from source code lack the
ability to provide idiomaticity in their use of the technology as well
as proper reproducibility safe-guards. In the first part, existing
approaches are formalized and discussed and a new approach is
introduced. The approaches are then evaluated using a suite of three
different examples.

This thesis provides a framework for formalizing operations involving a
layered file system, containers and images, and a novel approach to the
creation of images using utility containers and layer donning fulfilling
the idiomaticity and reproducibility criteria.

\chapter*{Zusamenfassung}\label{zusamenfassung}
\addcontentsline{toc}{chapter}{Zusamenfassung}

Container als Methode der Applikationsverteilung sind der neueste Trend
in der Softwareentwicklung. Sie gestatten es Entwicklern, ein
ausführbares Abbild zu erstellen, das die Anwendung mitsamt aller
Abhängigkeiten ent-hält. Dieses Abbild kann dann von Nutzern in einer
kontrollierten Umgebung mit virtualisierten Ressourcen ausgeführt
werden. Komplexe Arbeitsabläufe für unternehmenskritische Anwendung und
Forschungsumgebungen verlangen einen hohen Grad an Wiederholbarkeit, die
durch eindeutig identifizierbare Abbilder als Grundlage der Berechnung
erreicht werden können.

In dieser Thesis wird gezeigt, dass die verbreitetsten Ansätze, ein
solches Abbild von bereits existierender Software oder vom Quelltext zu
erstellen, die Möglichkeit vermissen lassen, Idiomazität in der
Verwendung der Technologie und echte Wiederholbarkeit zu gewährleisten.
In einem ersten Teil werden vorhandene Ansätze formalisiert und
diskutiert sowie ein neuer Ansatz vorgestellt. Anschließend werden die
Ansätze anhand einer Sammlung von Beispielprogrammen bewertet.

Diese Thesis bietet ein Framework zur Formalisierung von Vorgängen mit
einem geschichtetes Dateisystem, Containern und Abbildern und einen
neuen Ansatz für die Erstellung von Abbildern mit Utility Containern und
Layer Donning, der die Forderung nach Idiomatizität und Wiederholbarkeit
erfüllt.

\newpage

\tableofcontents

\listoffigures
\begingroup
\let\clearpage\relax
\listoftables
\endgroup

\chapter*{List of Abbreviations}\label{list-of-abbreviations}
\addcontentsline{toc}{chapter}{List of Abbreviations}

\setstretch{1.35}

\begin{description}
\tightlist
\item[API]
Application Program Interface
\item[AppC]
Application Container, a container format
\item[CD-ROM]
Compact Disc in the Read-Only-Memory variant
\item[CPU]
Central Processing Unit
\item[CWL]
Common Workflow Language
\item[DEB]
Debian package format
\item[DWG]
Data Working Group, a team under the umbrella of the GA4GH
\item[GA4GH]
Global Alliance for Genomics and Health
\item[GCC]
GNU Compiler Collection
\item[GNU]
GNU is not Unix
\item[HTTPS]
Hypertext Transfer Protocol over SSL, an encrypted and signed variant of
the Hypertext Transfer Protocol.
\item[ISO]
International Organization for Standardization
\item[JSON]
Java Script Object Notation
\item[JVM]
Java Virtual Machine
\item[NPM]
Node Package Manager, a code repository for NodeJS
\item[OCI]
Open Container Initiative
\item[PGP]
Pretty Good Privacy
\item[RPM]
Red Hat package format and package manager
\item[TSV]
Tab Separated Values
\item[VM]
Virtual Machine
\item[WDL]
Workflow Description Language
\end{description}

\setstretch{1}

\addtocontents{toc}{\protect\setcounter{tocdepth}{2}}

\chapter{Introduction}\label{introduction}

Software-aided research has specific requirements towards the employed
tools: It is important for meaningful research that results of one study
are reproducible by other researchers. The current approach of
publishing the source code only is often insufficient as important
details on the exact executed code are missing (e.g.~additional
libraries and compiler options/versions). A study by Collberg \emph{et
al.} about repeatability of studies backed by code came to the
conclusion that only for \(54\%\) of reviewed articles the code was
compilable at all, just \(32.3\%\) of studies were compilable in less
than \(30\) minutes \autocite{repro-software}.

At the same time, the industry is shifting away from so called big-bang
releases towards policies of continuous integration and deployment to
enable quick and timely feedback from users. Continuous deployment can
only reasonably be done when the compilation, testing and deployment
stages of the build process may execute without the need for a human
operator.

Software components usually rely on other software to compile and
sometimes to operate correctly. Different subsets of these dependencies
have to be available in the environment during the different stages of
the build mentioned. It is generally desirable to have exactly the same
versions of these dependencies during all stages on all involved
machines to decrease the chances of introducing unreproducible failures
caused by incompatible versions. Furthermore, fast build-and-test cycles
give valuable feedback to the developers as well as proving that the
software works before deploying it to a large scale production
environment.

The recently developed container concept popularized by
Docker®\footnote{Docker® is a registered trademark of Docker, Inc.}
promises to resolve this problem by encapsulating all dependencies
(software and other files) of an application into a single and
lightweight standalone executable archive. This provides other
researchers or end users with a `binary image in which all the software
has already been installed, configured and tested'
\autocite{Boettiger:2015:IDR:2723872.2723882}.

In addition to providing encapsulation it is necessary to describe the
exact steps to be taken to recreate a software image in executable form.
Ideally each time the build is executed bitwise equivalent images are
produced which enables verification of any images provided by authors.
However, current approaches to building software in containers make it
harder to ensure strong reproducibility.

There are various techniques that can be used to employ a container
system to reproducibly build software and to speed up the image creation
process. By creatively exploiting concepts and build environments like
squashing the resulting layers or designing a container for more than
one task it is possible to optimize different aspects, but using the
technology differently than the developers intended causes a reliance on
this particular implementation. There are guides for each popular
programming language and environment recommending the `correct' use of
the technology in patterns called `idioms' that ensure using the
language's or environment's features best. Some of the more popular
optimizations mentioned above can not be considered idiomatic with
regard to the officially endorsed best practices of container use.

The objective of this thesis is to develop a method for idiomatic and
reproducible software builds able to use existing software build tools
for automatic packaging. There is a selection of approaches already
available that facilitate the use of containers in software development
workflows. However, as will be shown in this thesis, neither of them
employ containers to their full potential.

The topic for the thesis was handed out by the bioinformatics group as
part of the ongoing Galaxy project which offers a platform for
`accessible, reproducible, and transparent computational biomedical
research.' \autocite{Goecks2010}. During the thesis the author was
employed by the Inxmail GmbH in Freiburg im Breisgau. The remainder of
this thesis is structured as follows:

Chapter \ref{theoretical-foundations} introduces technologies and
approaches to virtualization and in specific to containers. Furthermore,
it defines a mathematical notation for the image building process and
presents an in-depth model of the currently used container
implementations.

Chapter \ref{building-and-delivering-software} introduces different
existing approaches to building software using containers. Afterwards,
it proposes a new approach focussing on reusability and idiomaticity and
introduces an implementation of this approach. The approaches are
compared using an experimental evaluation and discussed.

Chapter \ref{application-mulled} gives an overview over the application
`Mulled' developed as part of the thesis. It shows the opportunities
offered by utilizing the approach introduced in chapter
\ref{building-and-delivering-software} and proves the applicability in a
real-world environment.

This thesis is based on the Docker Engine 1.10. In accordance with the
advisor this thesis was not rewritten for the newer version 1.11 which
was released just before submission. The major difference is the
adoption of the Open Container specification as the basis for Docker
containers (see section \ref{open-containers}).

\chapter{Theoretical Foundations}\label{theoretical-foundations}

This chapter introduces different concepts for virtualization and
explains their differences. Afterwards, different containerization
implementations and formats are presented. Finally, the Docker model and
software are introduced in greater detail.

\section{Virtualization}\label{virtualization}

Virtualization as a concept dates back to the first mainframes and the
sharing of resources via time-split approaches. Innovations in the
separation of processes belonging to different users at the operating
systems level allowed shared usage of expensive hardware, thereby giving
software the `illusion that it has exclusive access to the underlying
hardware platform' \autocite{p34-crosby}. This illusion can be used to
separate processes both for increased security (processes can't
influence each other) and for simpler process execution models
(processes can assume full control over the machine).

Support for virtualization on modern processors made it possible to not
only separate processes efficiently, but also to run different operating
systems and even emulate other hardware platforms. Different solutions
provide virtualization at different layers in the soft- and hardware
stack with extremely different characteristics and areas of application.
Exemplary implementations are discussed in the following sections.

\subsection{Full Operating System
Virtualization}\label{full-operating-system-virtualization}

Multiple operating systems can be executed on one physical host by
different software tools. One such software is QEMU\footnote{\url{http://qemu.org/}}
which is able to emulate a whole computer by providing software
implementations for all peripherals normally found in a personal
computer, including hard drives and network cards. If the architecture
of the virtual computer does not match the architecture of the hosting
machine the processor is emulated in software as well. However, if the
architecture matches, QEMU is able to execute code directly on the host
CPU (see the introduction by Bartholomew \autocite{bartholomew2006qemu})
using a technology called KVM\footnote{\url{http://www.linux-kvm.org/}}
that allows programs to control a virtualization extension on the CPU.

By fully emulating a computer QEMU is able to execute multiple (possibly
different) operating system instances independent of each other on one
host. Each instance however incurs a cost in the form of the memory and
processing time needed to run the systems. If similar operating systems
are used core services are repeated in each instance, leading to
inefficient resource usage.

\subsection{Kernel based
Virtualization}\label{kernel-based-virtualization}

It is often not necessary to have completely separate operating system
instances for each server. Using one kernel instance for multiple
virtual machines usually requires less memory and fewer CPU cycles since
internal maintenance tasks are only executed once.

It is vital for the security and for the safety of a process that only
interactions specifically allowed by the developer between this and
other processes can take place, for example using Shared Memory or other
general Inter-Process Communications. Accordingly, it is one of the core
tasks of an operating system to separate and protect processes from each
other. The operating system (with possibly the help of the underlying
chipset) uses different techniques to make it impossible for normal
unprivileged processes to access other processes memory representing the
state of the process, for example using virtual memory addressing.

There has been a way to change a process' view of the file system since
Unix Version 7, namely the \lstinline!chroot()! system call. It allowed
a system administrator to change the meaning of the root directory
(`\lstinline!/!') for a process. Initially there was no security code in
place to make an escape from this isolation impossible. According to
Siebenmann (see \autocite{chrootHistory}), this probably wasn't the
intention initially anyway, it was intended to be used to provide an
application with its desired environment. In more recent development
iterations it was made impossible for processes running not in the
context of the super user account to escape the directory they are
confined in which enables its use in security related measures.

Recently, the concept of \emph{namespaces} was introduced into the Linux
kernel, allowing thorough isolation of \emph{process groups} (a process
and possible descendant processes) from the `outside' world. Each
process group `adds an additional indirection {[}\ldots{}{]} to the
naming/visibility of some Unix resource space'
\autocite{addingGenericContainers} which can be used to effectively
simulate a different machine to the running process. This means that a
process encapsulated in a namespace has no access at all on other
processes, files, network interfaces etc. outside of its namespace since
the operating systems makes it impossible to even perceive their
existence. Processes outside of any namespace, for instance interactive
console sessions, have an unrestricted view of the whole system,
including the namespaced processes. This is an implementation of the
\emph{one-way isolation} concept as defined by Liu et al (see
\autocite{liu2000intrusion}). It is also an instance of an `API layer
VM' as described by Yan Wen \emph{et al.} in their comparison of
virtualization technologies focussing on untrusted code execution
\autocite{surveyVirtualization}.

In contrast to the virtualization methods described above, the overhead
incurred by namespace isolation is almost negligible according to the
performance comparison by Felter \emph{et al.} (see
\autocite{updatedPerformanceComparison}). Compared to systems like Xen
Soltesz \emph{et al.} estimate the performance advantage at `up to 2x
{[}\ldots{}{]} for server-type workloads and {[}they{]} scale further
while preserving performance' \autocite{p275-soltesz}. Due to the
control over all relevant Unix resources it is however possible to
achieve a level of separation between processes comparable to a complete
operating system virtualization, except of course the sharing of the
underlying kernel. From the perspective of an isolated process there is
no difference in behaviour of the system.

Related to but not part of namespaces is the implementation of
\emph{control groups} in the Linux kernel. They `provide a mechanism for
aggregating/partitioning sets of tasks {[}\ldots{}{]} into hierarchical
groups {[}\ldots{}{]}' \autocite{cgroups}. With this technology, an
operator can impose resource limits on a process group formed by a
process and its descendants. From the isolated process' view the machine
it is running on only has the defined amount of memory, CPU power etc.
The implementation allows flexible specifications of limits by
classifying tasks, and imposing constraints based on the membership of
processes into these classes. A management service can manage such
control groups independently of the processes' implementation.

Another implementation of a similar technology are Zones provided by the
Solaris\footnote{\url{https://www.oracle.com/solaris/}} operating
system. A dedicated `global' Zone is able to manage a set of dependent
Zones, each suited for `the needs of {[}the{]} particular applications'
\autocite{solaris-zones}. The small market share of Solaris on servers
and most importantly on developer machines however prohibited a
widespread adoption of the technology.

Since the kernel is shared between process groups any vulnerability in
the kernel can potentially be exploited to allow a process to break out
of its namespace and gain full unrestricted access to the host system.
There has already been a number of attacks against the containment
implementations in recent years, for example the exploitation of
unneeded, but retained Linux capabilities Traditionally, a process in
UNIX has access to system-wide permissions if it has the effective user
id \(0\), which is reserved for the \lstinline!root! user. Capabilities
`divide{[}s{]} the privileges into distinct units'
\autocite{manCapabilities} which can be held individually. See the
presentation by Barth and Luft \autocite{dockerAndSecurity} for details
of this attack. In this report on the state of container security, the
authors add a warning that any privileged process which can connect to
some kernel modules or can access certain system files could potentially
break out of its container. Implementors have fixed a number of
vulnerabilities already (e.g.~CVE-2014-3499 {[}local privilege
escalation{]} and CVE-2015-3629 {[}namespace breakout{]}), but due to
the high number of system calls in the Linux kernel which are
potentially vulnerable an exploit is always a possibility. Additional
measures to secure the systems against a breakout are recommended.

The `CIS Docker Benchmark' by Goyal \emph{et al.}
\autocite{cis-docker-benchmark} provides a reference for a fast and
secure environment using the Docker software. It contains detailed
recommendations for running Docker containers, including enabling
additional security features on the host machine.

\subsection{Intra-Process
Virtualizations}\label{intra-process-virtualizations}

Application servers such as GlassFish\footnote{\url{https://glassfish.java.net/}}
provide an environment for individual applications that hides the
presence of instances of other applications inside the same process.
Similar to cgroups the operator can set restrictions on resource usage.
It is usually possible to start and stop applications running inside the
server without restarting the host process.

The host process (the application server) provides services to the guest
using environment-specific interfaces. This limits the choice of
possible development languages to the ones supported by the application
server. For instance, the GlassFish server supports execution of
programs written in a Java Virtual Machine (JVM)-based language, such as
Java or Groovy.

In general, multiple threads of one process can be considered to be a
form of virtualization as well. From the point of view of each thread it
has full control of the computer each time it is running, and direct
communication with other threads (even in the same process) can only
happen using interfaces the runtime environment exposes to them.

The separation between components in this virtualization model isn't
impermeable because of the inherent coupling between the different parts
of the process. Cross-component memory access for example is neither
preventable by the environment nor is it desirable to do so as the
components necessarily share an address space and an operating system
process and may need the communication channel provided by the shared
memory to accomplish their task. If one of the threads or applications
however causes a process failure, the entire application server crashes.
Additionally, the restriction of environment to compatible languages
greatly limits the choice of technologies for new projects that have to
be executed within the existing context.

\subsection{Summary}\label{summary}

The motivation for developers of containerization software is to create
an environment that supports a wide range of runtime environments to be
able to select the right programming language, frameworks etc. for each
individual application. On the other hand, this environment should incur
minimal overhead in terms of performance and resource requirements.

There are different methods for the virtualization of resources working
on different level in the system and separate components differently.
Namespaces in combination with cgroups provide a `lightweight'
\autocite{linux-journal-merkel-dirk} virtualization technology with
negligible performance impacts if it is possible to share a kernel
instance among different applications. Other virtualization technologies
either virtualize a whole operating system with the associated cost or
require a close coupling between different applications.

\section{Container Formats and
Implementations}\label{container-formats-and-implementations}

Containers are a loose-defined concept for encapsulation of applications
and their execution. They are usually implemented with namespaces and
cgroups, but implementations using other technologies are in their early
stages and expected to be released in the near future.

There are multiple different and incompatible formats by different
vendors. In this section some of them are introduced.

\subsection{Docker Images}\label{docker-images}

The Docker platform amounts to a market share of almost 80\% of
respondents to a survey conducted by O'Reilly and Ruxit (see
\autocite{stateContainers}) in 2015. According to Peter Biggar (see
\autocite{containerWars}), the term `container' has been synonymously
used with `Docker' until recently.

Docker published an image specification for images of version \(1\)
\autocite{docker-image-spec-v1}. Newer versions of the Docker Engine
(starting with version 1.10) can still import such images, but convert
them internally into a newer, not publicly specified image format. This
format will be discussed in greater detail in section
\ref{docker-engine}.

\subsection{App Containers}\label{app-containers}

Because of disapproval towards the company policy of Docker Inc. CoreOS
Inc.\footnote{\url{https://coreos.com}} has been developing its own
container engine \emph{rkt} that fulfils the design goals CoreOS expects
from a production-ready container engine (`CoreOS is building a
container runtime, rkt', \autocite{rocketAnnouncement}). Their container
engine is built around a specification for containers called `App
Container' (see the specification \autocite{appcSpec}).

The design goals CoreOS requires from a production-ready container
engine as mentioned in the announcement are:

\begin{description}
\tightlist
\item[Composability]
This is the ability to build higher-order tools from small building
blocks. According to CoreOS, the architecture of Docker consists of too
tightly coupled interfaces that inhibit clear separation of concerns and
make composing additional tools unecessarily hard. The monolithic design
of the Docker daemon is considered to prevent easy addition of features
by external developers.
\item[Security]
Cryptographic securities and isolation principles should be present
`from day one' \autocite{rocketAnnouncement}. The first App Container
release already allowed referencing containers by uniquely identifying
cryptographic hashes and included instructions for crytographic
signatures.
\item[Image Distribution]
The image distribution model of Docker is built around Docker Hub.
Private and independent registries used to use a different protocol than
Docker Hub, and still have to use an additional prefix (the address of
the registry). For App Containers, the specification comittee designed a
federated and decentralized discovery protocol without mandating the
underlying transport protocol. This makes independent developments with
other transmission protocols possible.
\item[Open]
Although the Docker software is open source the ecosystem is not as open
as the CoreOS developers require it to be. The interfaces powering the
tools should be developed as an open specification by an open community
of developers and not a private company.
\end{description}

The proposed specification of the App Container format is independent
from the rkt engine, and rkt is just one of a growing set of
implementations. The specification is licensed under the Apache license
(see \autocite{appcSpec}). It is structured into four parts: A
specification for images, for image discovery, for a concept called
`pods' and for an executor.

An App Container image consists of the (flat) root file system
represented by a directory and a manifest file describing metadata of
the image, packed together in an archive file. This model differs
greatly from the Docker specification which allows and encourages
multiple layers of images stacked on top of each other.

Each image is identified by a cryptographic hash of its contents. There
are also provisions for encrypting and signing the images using Pretty
Good Privacy (PGP). Execution restrictions can be specified in a
metadata block: It allows to limit the resource usage (memory, CPU etc.)
as well as which Linux capabilities the process may retain.

Names can be assigned to images, conventionally of the form
\lstinline!domain.com/name!. The App Container Specification describes a
decentralized way to resolve such names using special instructions in a
file retrievable from that name using HTTPS. These instructions also
contain a cryptographic signature of the image, which enables forwarding
the trust placed into the HTTPS connection towards the actual app
container image.

A complete executable unit defined in the specification is called a
`pod'. It specifies a set of containers to be executed in parallel and
in the same namespace and cgroup. This means that containers in the same
pod are able to communicate with each other, for example using process
signals.

The executor specification defines the exact environment a container can
expect when it is instantiated. The way the abstract metadata and
environment constraints from the pod and image manifests are realized
prior the execution is defined by the specification.

\subsection{Open Containers}\label{open-containers}

In an effort to create an industry standard for containers and their
environment the Open Container Initiative (OCI) under the umbrella of
the Linux Foundation was founded. The goals of the container format with
the same name are, as postulated in their charter, `to
build{[}\ldots{}{]} a vendor-neutral, portable and open specification
and runtime' \autocite{charterOCI}. As a start, the new specification is
based on the format used by Docker.

The foundation's so called `Technical Developer Community' is tasked
with harmonizing the OCI specification with the App Container
specification (see \autocite{charterOCI}). Some of the developers behind
the App Container specification are members of this community. Judging
by the number and size of the companies backing this format it is to be
expected that the adoption of the specification will increase in the
near future.

\subsection{Summary}\label{summary-1}

All introduced container formats are quite young and all of them are
still in development. It is probable that the fragmentation will be
resolved and a common standard emerges. The `open containers' format
already tries to bring adaptors of different standards together to
create a shared specification. Although solutions exist that create a
bridge between containers of different specifications, it is in the
interest of all adaptors to establish one format with possibly multiple
implementations. The format with currently the most widespread adoption
is the format in use by the Docker engine.

\section{Docker Engine}\label{docker-engine}

The Docker software uses the support in the Linux kernel for isolated
containers, and supplies a toolchain and ecosystem to build, execute,
manage and share such containers. Although commonly obscured there is a
difference between a container and an image: An image serves as an
abstract description of a container, comparable to the notion of classes
and objects in common object-oriented programming languages.

At the time of writing, the Docker ecosystem was in the process of
changing the layer addressing system from random identifiers to a
content addressable scheme. This model doesn't change the semantics of
file system accesses; only the way of addressing the layer differs. The
consequences of this shift will be discussed below.

\subsection{Layered File System}\label{layered-file-system}

This section introduces the concepts defined by the Docker ecosystem for
accessing, writing and deleting files in a layered file system. A
mathematical formulation is used to make comparing different methods for
building such images possible. The concepts described here are
documented from an implementation point of view in the `Docker Image
Specification' \autocite{docker-image-spec-v1}.

For the following definition the concept of a \emph{directory} is
needed. This definition closely follows the semantics that are encoded
in common file systems. It allows the lookup of a file by name, which is
an instance of a \(\mathtt{String}\).

\begin{definition} A partial function \(d\) of the form
\(d : \mathtt{String} \nrightarrow \mathbb{F}\) is called a
\emph{directory}, where \(\mathbb{F}\) denotes the set of
files\footnote{Symbolic links are as files as well, they are treated as
  if their target is stored in a file with the same name.}.\end{definition}

A file name may contain directory names, such as
\lstinline!/tmp/data/test.txt!. A file is contained in a directory if
its file name starts with the name of the directory ending in a slash.
The file mentioned above is contained in the directories
\lstinline!/tmp/! and \lstinline!/tmp/data/! as well as in the root
directory \lstinline!/!.

\begin{definition} A file with name \(f\) is contained in a directory
with name \(p\) ending in a slash if and only if \(f \sqsubseteq p\).
Its name can then be expressed by the concatenation of \(p\) and a
suffix \(\tilde{f}\), such that \(f = (p)(\tilde{f})\).\end{definition}

During execution a container may have access to multiple directories
`mounted' into different location in its file system. These directories
can be provided by other containers for cross-container file sharing,
but also directly from the host system, for instance to enable
persistent data storage. For example, if the directory \(d^\prime\) is
mounted into the path \lstinline!/data! of the directory \(d\), all
accesses to \(d\)\lstinline!/data! and files contained in it are
evaluated with \(d^\prime\) without the prefix. In this example the file
with name \(d\)\lstinline!/data/test/a.txt! is given by
\(d^\prime\left(\mathrm{/test/a.txt}\right)\).

\begin{definition} The result of \emph{mounting} the directory
\(d^\prime\) at the location \(p\) into the directory \(d\) is given by
\[\hat{d}\left(f\right) = \begin{cases}
  d^\prime\left(f^\prime\right) & f = (p) (f^\prime)\\
  d(f) & \mathrm{otherwise}\\
\end{cases}\]\end{definition}

\subsubsection{Image Layer}\label{image-layer}

The smallest entity in the Docker ecosystem is an \emph{image layer}. It
represents the \emph{changes} made to a directory, i.e.~it is the
difference between two versions of the same directory, and also contains
meta data that further describe these changes.

\begin{definition} A change is either the creation of a file or its
deletion. The set of changes is defined by
\(\mathcal{C} = \mathbb{F} \cup \left\{\bot\right\}\), where \(\bot\)
denotes the deletion of a file. A layer is a partial mapping
\(l \in L : \mathtt{String} \nrightarrow \mathcal{C}\). The argument of
\(l\) is the name of the file that was changed.\end{definition}

The meta data of a layer are serialized on disk in the Java Script
Object Notation (JSON), as defined by Bray and Crockford in RFC 7159
\autocite{jsonRFC}. Information stored in the meta data is used to
construct the image layer stacks and to start containers.

\begin{definition} There is a function
\(meta_l : \mathtt{String} \nrightarrow \mathtt{JSO}\) with \(l \in L\)
and \(JSO\) the set of instances of the data types defined by
JSON.\end{definition}

We refer to image layers (and later to containers as well) by their IDs.
They are usually generated randomly.

\begin{definition} An ID is an 256-bit wide number. It is represented in
hexadecimal encoding.\end{definition}

To uniquely identify an image, the value of the meta data key \emph{id}
is restricted to appear at most once in any given system. It is
accordingly possible to get a layer with this ID:

\begin{definition} There is a partial function
\(layer : \mathtt{String} \nrightarrow L\) such that
\(layer\left(meta_l \left('\mathtt{id}'\right)\right) = l\) and
\(i, j \in \text{ID}, \forall i: \exists layer(j) \Leftrightarrow layer(i) = layer(j)\).\end{definition}

\subsubsection{Image Layer Stacks}\label{image-layer-stacks}

Multiple layers can be stacked on top of each other providing a unified
view over a list of file changes. In Docker images of version 1, each
layer has either one parent layer, or no parent layer (see below for
version 2). This relationship is expressed in the meta data: The key
\emph{parent} is either the ID of the parent image layer, or it is set
to \lstinline!null!. The resulting structure is a set of unidirectional,
acyclic graphs of image layers, and these trees are (as can be seen
directly) disjoint, and therefore form a \emph{forest}.

\begin{definition} The relation \(P \subset L \times L\) \footnote{It is
  easy to see that \(P\) is a proper subset: In any
  \(L \times L \ne \varnothing\) there exists at least one element
  \(\tilde{l}\) such that \(\tilde{l}\) has no parent.} denotes the
parent relationship. A pair of layers \((i, p)\) is element of \(P\) if
and only if \(\mathtt{parent}(i) = p \Leftrightarrow\)
\(meta_i\left('\mathtt{parent}'\right) = meta_p\left('\mathtt{id}'\right)\).\end{definition}

For easier reference image layers can be given a \emph{tag}. These tags
are composed of a \emph{repository name} and a \emph{version}, separated
by a colon (\lstinline!:!). By convention, the version
\lstinline!latest! is the least recently published version. Apart from
this the semantics of the version are left to the publisher. All image
layers tagged with a common repository name, but a different version
form a shared \emph{repository}.

\begin{definition} The function
\(layer_T : (\mathtt{String} \times \mathtt{String}) \nrightarrow L\)
maps a repository name and a version to an image layer.\end{definition}

To execute a container based on an image layer it is necessary to have
all image layers available that are reachable by the parent relation
\(P\). These image layers forming the transitive closure of \(P\) are
\emph{included} in the tag.

\begin{figure}[htbp]
\centering
\includegraphics{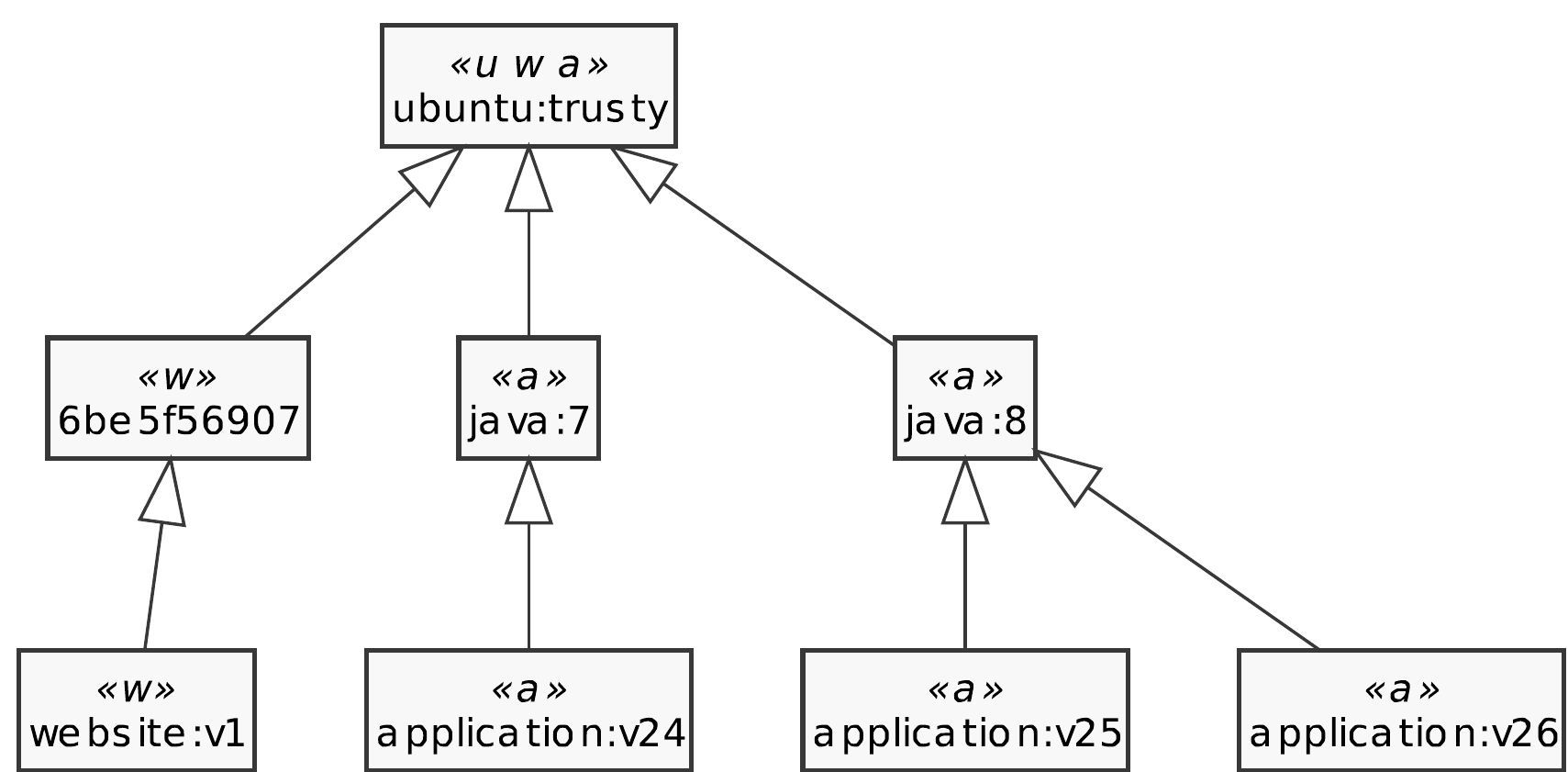}
\caption{\label{fig:docker-layers}Docker image layers}
\end{figure}

Figure \ref{fig:docker-layers} shows the structure formed by a set of
image layers. The composition arrows point to the parent layer, while
the \lstinline!ubuntu:trusty! image layer at the top has no ancestors.
Above the tag of an image layer the repositories in which an image layer
is included are marked with an abbreviation. The application image (`a')
is based on the Java image which in turn is based on the Ubuntu image
(`u'). The two versions 25 and 26 of the application have been chosen to
run with the Java runtime in version 8 and share this image, while the
older application version 24 still uses the runtime in version 7. The
website image (`w') uses an untagged image layer that is in turn a child
of the Ubuntu image (`u').

A stack of image layers forms an abstract specification for a concrete
container. An instantiated container sees a combined representation of
this stack, as patented by Hipp \emph{et al.} in `Method and system for
an overlay filesystem' \autocite{ofs-patent}. Containers based on the
application image of version 25 therefore has access to the files
`changed' in the layers \lstinline!application:v25!, \lstinline!java:8!
and \lstinline!ubuntu:trusty! according to the following definition:

\begin{definition} Let \(c\) be a container instantiated from image
\(i\). Then (for processes in the container) the file with file name
\(f\) is given by the function \[\mathtt{file}_i(f) = \begin{cases}
    \delta                                 & \delta = i(f)\\
    \mathtt{file}_{\mathtt{parent}(i)} (f) & (i, p) \in P\\
    \bot                                   & \mathrm{otherwise}\\
\end{cases}\]\end{definition}

\subsubsection{Content Addressable Image
Layers}\label{content-addressable-image-layers}

The identifiers of image layers were generated randomly by older
versions\footnote{until and including version 1.9} of the Docker
software. It was possible to have the same image layer with multiple
identifiers, but more importantly there was no intrinsic connection
between the identifier and the contents. This made testing transferred
image contents for any modifications during transfer impossible.

To mitigate this risk Docker introduced content addressable image
layers. Similar to the addressing scheme used by the Git version control
system, layers are identified by a cryptographic hash of their contents.
This makes it impossible to have accidental or intentional modifications
during transit as long as the correct hash is known by the receiving
party.

In the content addressability universe, an image is described by a
\emph{manifest} denoting the order of layers identified by their hash
that in combination constitute the full image and the configuration
object that is used to instantiate the container. It is identified by a
cryptographic hash of its contents and contains the combined
configuration data from the composing layers, using the semantics from
above.

In contrast to the legacy addressing model the layers do not form a tree
and layers do not `know' their parent. Using the notation from above, it
can be formalized as follows:

\begin{definition} The sequence of cryptographic hashes
\(h_{\tilde{c_{i^2}}} = \left(h_1, h_2, \dots\right)\) stored in the
configuration object \(\tilde{c_i}\) denotes the stack of image layers
composing the image named by \(\tilde{i^2}\). A cryptographic hash is an
instance of \(\mathtt{Hash}\).\end{definition}

The corresponding layer can be found in the image store using the
cryptographic hash.

\begin{definition} There is a partial function
\(layer^2 : \mathtt{Hash} \nrightarrow L\) such that
\(layer^2\left(h\right) = l\) and a total function
\(hash : L \rightarrow \mathtt{Hash}\) \(hash(l) = h\).\end{definition}

The \(layer^2\) function is partial because not every cryptographic hash
has a known image layer associated with it. However, it is always
possible to calculate a hash for an image layer.

If an image in the version 2 format is instantiated, the resulting
container uses a file name resolution scheme similar to the one from
version 1. However, instead of recursively using the parent-child
relationship it uses the order imposed by the configuration object of
the image:

\begin{definition} For processes in a container instantiated from image
\(i^2\) in version 2 format the file with file name \(f\) is given by
the function \(\mathrm{file}\), where \(h = h_{\tilde{c_{i^2}}}\),
i.e.~the sequence of layers in the image:\end{definition}

\[\mathtt{file}_{i^2}(f) = \begin{cases}
    \delta                                 & \delta = layer\left(\left(h\right)_1\right)\left(f\right)\\
    \delta                                 & \delta = layer\left(\left(h\right)_2\right)\left(f\right)\\
  & \vdots \\
    \delta                                 & \delta = layer\left(\left(h\right)_n\right)\left(f\right)\\
    \bot                                   & \mathrm{otherwise}\\
\end{cases}\]

As long as no layer contains either a deletion or an addition, the
layers are tried in order until a match is found. If the file was never
`changed', it is considered to be non-existent.

If the user is able to trust the validity of the manifest, for example
because it was retrieved over a secure channel from a trusted source, it
is possible to validate layer files retrieved via an insecure connection
against these checksums. The trust placed in the manifest is expandable
to the layer files due to the use of cryptographic hashes.

\begin{figure}[htbp]
\centering
\includegraphics{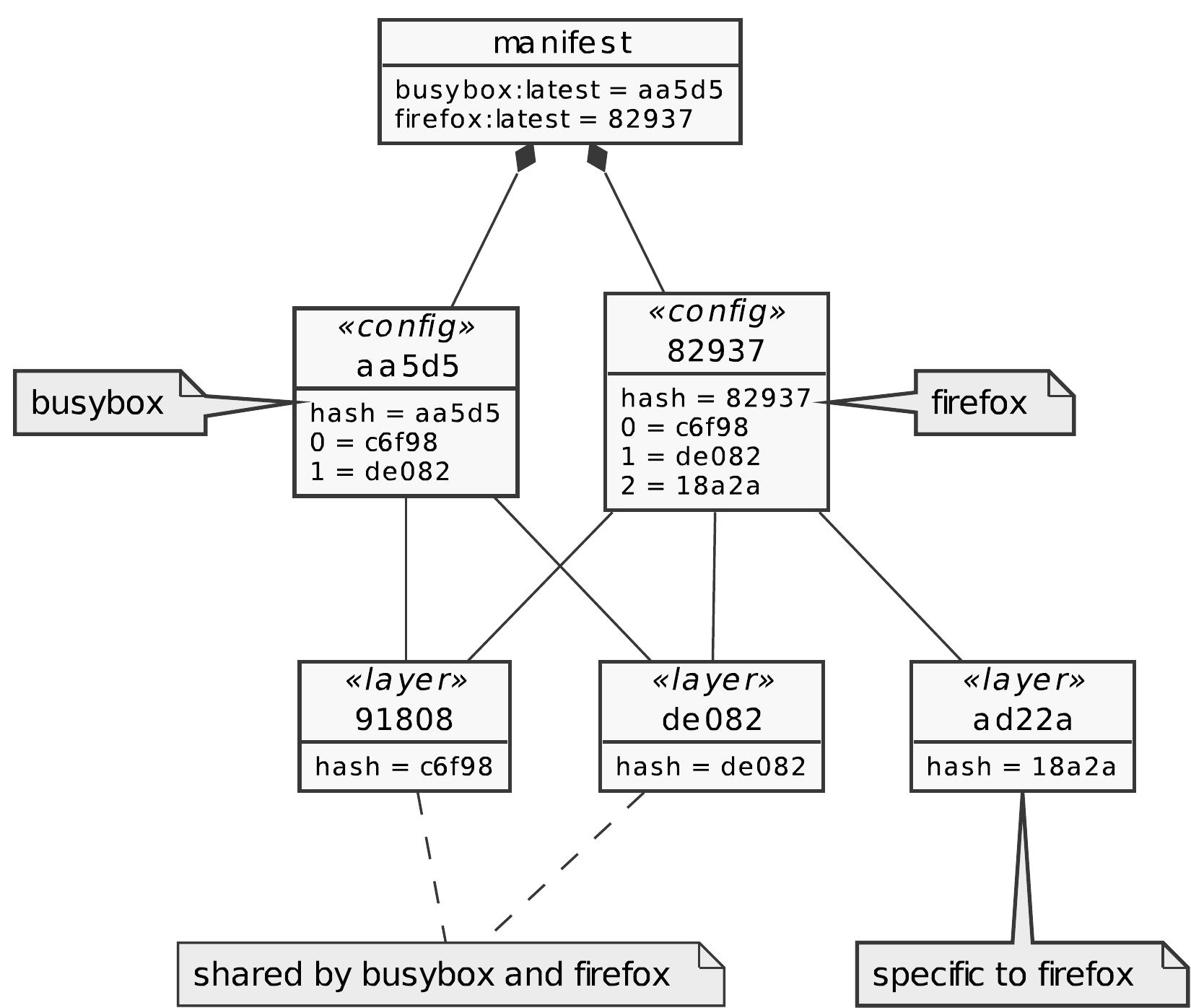}
\caption{\label{fig:docker-content-addressable-layers}Content addressable image layers}
\end{figure}

Figure \ref{fig:docker-content-addressable-layers} shows an example
containing two images with content addressable layers: The latest
\lstinline!busybox! image (using the configuration object
\lstinline!aa5d5!) as well as a \lstinline!firefox! image (using the
object \lstinline!82937!). The latter is based on the former, i.e.~it
adds a new image layer on top of the existing \lstinline!busybox! stack.
The manifest specifies the repository names and tags for the two images
to point to a configuration object for each of the images and to the
cryptographic hashes of the layers in the correct order. The
\lstinline!firefox! image shares the two of the \lstinline!busybox!
image, and adds another on top.

\subsection{Provisioning with
Dockerfiles}\label{provisioning-with-dockerfiles}

\begin{lstlisting}[language=Dockerfile, caption=Exemplary Dockerfile, float=btph, label=lst:example-dockerfile]
FROM ubuntu:trusty

RUN apt-get update &&\
    apt-get install -y netcat-traditional

EXPOSE 12345
USER nobody

CMD ["/bin/bash", "-c", \
 "while true; do nc.traditional\
 -e /bin/cat -klp 12345; done"]
\end{lstlisting}

Dockerfiles are an approach built into the Docker software components to
create images based on instructions in a text file. A Dockerfile
specifies a base image and any subsequent changes to make to that image.
Each change is recorded in an image layer, and stacked on top of the
previous change. A subset of the available commands is introduced in the
following, based on the official documentation\footnote{\url{https://docs.docker.com/reference/builder/}}:

\begin{description}
\tightlist
\item[FROM]
Sets the base image for the following commands. This command is
mandatory and has to be first in the file.
\item[RUN]
Executes a command in the context of a \emph{container} based on the
current image. The resulting image is constructed from the state of the
container after the command terminates.
\item[EXPOSE]
Containers may choose to expose certain ports to their host computer to
offer network services such as web servers. All subsequent layers will
carry this meta information.
\item[USER]
Each command in the Dockerfile is executed within the context of a user.
By default this is the supervisor user \lstinline!root! which is allowed
to do everything in the virtualized system. Commands in layers below
this command only have the permissions of this user, but later
\lstinline!USER! invocations can switch back to \lstinline!root!.
\item[ADD/COPY]
With these commands, external files can be copied into the container at
arbitrary locations. \lstinline!COPY! is restricted to files in the
directory containing the Dockerfile, while \lstinline!ADD! is able to
download files from an URL and to unpack files into a directory in the
container.
\item[CMD]
Registers the argument as default command for this image. If no command
is specified when a container is instantiated, this will be used.
\end{description}

The listing \ref{lst:example-dockerfile} shows an example for a simple
containerized echo service. The \lstinline!netcat! tool is installed
with Apt\footnote{\url{https://wiki.debian.org/Apt}}, and is configured
to run as the user \lstinline!nobody! on container start. When a
connection is made to the TCP port 12345 it will start
\lstinline!/bin/cat! to handle this connection, giving the impression of
an echo server. This port is exposed to the outside of the container,
giving other containers the possibility to use this service.

Dockerfiles are executed by the Docker daemon by alternately creating
containers and images, each time applying one change corresponding to
one line in the file in an image layer. As an optimisation, Docker
reuses intermediate images generated by earlier stages if the commands
have not changed. For instance, if line \(7\) is changed in the example
listing \ref{lst:example-dockerfile}, Docker uses the image generated by
line \(6\) and applies the changes from line \(7\) and \(9\) only.

\chapter{Building and Delivering
Software}\label{building-and-delivering-software}

In this chapter different approaches to deliver software are outlined
and evaluated. The focus is set on a containerized build deployment
environment. Firstly, the process of building software is formalized and
then used to investigate the impact the approaches have on the final
deliverable.

\section{Introduction}\label{introduction-1}

For the purpose of this evaluation the meaning of `building software'
has to be defined exactly. While the notion of building software is
usually understood to refer to the generation of executable programs in
binary form from a set of source code files, it is also used in a much
wider sense: Generation of documentation, minification of programs in
interpreted languages, and the execution of test suites are often
included into this process. This process typically consists out of
multiple steps that have to be executed in order, and after the final
step has completed a deliverable artefact is present.

Ideally, an artefact is the result of a \emph{pure} function of the
source files. Environmental influences are not allowed to take place in
this model, for example the current date and time should not influence
the outcome. Current build systems however do not fully implement this
model, but there are attempts to rectify this, for example in the Debian
community\footnote{\url{https://wiki.debian.org/ReproducibleBuilds}.}.
Since environmental influences usually manifest themselves in comparably
small binary changes and don't change the behaviour of the system at
large this thesis only considers the source files as input for a builder
function which is defined as:

\begin{definition} A function \(f_\xi\) of the form
\(f_\xi : D \rightarrow D\) is a \emph{builder} in the environment
\(\xi\), where \(D\) is the set of directories as defined
above.\end{definition}

The difference between two directories, i.e.~the changes that transform
a directory \(a\) into directory \(b\) is the inverse function of a
builder. It is defined as follows:

\begin{definition} The function
\(\mathrm{diff} : D \times D \rightarrow L\) gives the difference
between two directories \(a\) and \(b\) defined as
follows:\end{definition}

\(\mathrm{diff}\left(a, b\right) = \tilde{i}\) with
\[\tilde{i}\left(x\right) = \begin{cases}
  b(x) & x \in \mathrm{dom}(a) \land \mathrm{dom}(b) \land a(x) \ne b(x)\\
  b(x) & x \notin \mathrm{dom}(a) \land x \in \mathrm{dom}(b)\\
  \bot & x \in \mathrm{dom}(a) \land x \notin \mathrm{dom}(b)\\
\end{cases}\]

\section{\texorpdfstring{`Legacy'
Approaches}{Legacy Approaches}}\label{legacy-approaches}

A containerized approach to software delivery is a relatively new
concept. Other technologies are deployed in a large number of systems.
This section will show situations and circumstances in which software
delivery with containers provides advantages, but also shows possible
downsides.

\subsection{Tarball / Installer}\label{tarball-installer}

Traditionally, much of the available open software was (and still is)
distributed as tarballs, as recommended by the GNU Coding Standards by
Stallman \emph{et al.} \autocite{gnuCodingStandards} Such distribution
tarballs contain all source files needed to compile the software with a
compiler installed on the users machine, producing the `non-source
files' such as executable programs.

There are conventions regarding the exact structure and dependencies of
tarballs for successful compilation, but package maintainers are not
bound by restrictions. It is therefore quite common to have special
cases for different packages. Additionally, a quite diverse set of
conventions exist. Examples for tool-supported package conventions are
Autotools or Cmake.

The most obvious advantage of this model is the direct availability of
the source code for each package, and the level of control the
maintainers have over distribution: They can just publish a new archive
file on their website, and any user can download and build it. For
convenience, Linux distributions like Gentoo\footnote{\url{https://gentoo.org/}}
and its derivates provide their users with tools and scripts to build
software from the source code published by its maintainers on their own
machines. The Homebrew\footnote{\url{http://brew.sh}} package manager
offers portable scripts (`recipes') to fetch, compile and install
software as well.

However, each user has to have a full compiler suite available to use
the software, and it takes a non-trivial amount of time to build complex
software from sources. The non-standard behaviour of many build systems
prohibits fast and seamless usage of software by inexperienced users.

Dependency management is left to the distributor of the software.
Versioning of shared libraries is not trivial, which makes the
management of libraries hard. Reproducibility is difficult to achieve as
the exact compiler versions and runtime environment has to be
reproducible as well.

Distribution of closed source software is usually accomplished with
installation programs that copy the results of the build process into
appropriate places in the file system and establish links to executable
programs. Installation programs are often available on CDs or
downloadable from the vendor's website.

\subsection{Package Repository}\label{package-repository}

It is common for Linux distributions such as Debian to be based on a
\emph{package} system. The life cycles of small units of functionality
(files, programs, metadata etc.), so called packages, are controlled by
a dedicated \emph{package manager}, as described by Mancinelli \emph{et
al}. \autocite{complexityRepos} Usually, these packages contain the
resulting binaries of a build process, executed on machines of an entity
the user trusts. The most popular and widespread package systems are
considered to be the RedHat Package Manager (RPM) and the Debian
packaging manager using the DEB format (see Mancinelli \emph{et al.}
\autocite{complexityRepos}), but there is a multitude of package
managers addressing different kinds of packages. Other instances are
Conda\footnote{\url{http://conda.pydata.org/}} designed for Python
packages and the Alpine\footnote{\url{http://alpinelinux.org/}} package
manager \lstinline!apk! for the Linux distribution with the same name.

To facilitate installation procedures package resolvers have been
implemented on top of package managers. They are responsible to resolve
any dependencies and/or conflicts of packages. Package resolvers load
packages and meta data from \emph{repositories}.

From a user perspective, this provides a simple way to install software
to a computer. Dependencies are handled automatically by the package
resolvers, so software with a deep dependency graph is installable as
easily as one without dependencies. Since most distributions ship with a
package resolver preinstalled usually no additional software is needed
to access all functionality in the packages stored in the repositories.

Package managers can ensure the integrity of the package with
cryptographic checksums, proving that no changes were made after the
build process completed on the build machine. It is left to the
maintainers of the build cluster to make sure that the sources and the
binaries are not tampered with before packaging it.

Usually, a package manager installs all packages into a shared area in
the file system, which makes it possible to reuse common dependencies,
for example the default libraries for executables. However, this is only
possible if the dependencies are `compatible' (as defined by the
dependency type, for example the same Application Binary Interface for
shared libraries on Linux).

Without additional steps, installed packages share a namespace in the
file system, i.e.~they are able to read and possibly write each others
files. They also share an execution namespace, which means they can see
and influence processes of other packages.

\subsection{App Stores}\label{app-stores}

A new model for software distribution was popularized in recent years
starting on smartphones, but now also on desktop computers: App Stores
like the Windows Store distribute platform specific binary packages that
contain all dependencies (apart from platform level interfaces) with
digital signatures, and provide `a greater degree of separation between
apps than {[}\ldots{}{]} traditional desktop apps'
\autocite{deliveringMetroApps}. Other popular app stores and
environments provide similar encapsulation methods: Android uses the
`underlying Linux security model, based on user IDs'
\autocite{applicationSecurityAndroid} to separate file ownership and
permissions, and iOS ensures that `apps run in a sandbox and have
limited interactions with other apps' \autocite{iosAppArchitecture}.

Users and developers have the advantage that all dependencies are
bundled together in one package. Each app can rely on its dependencies
being there. The platform controls the exact details of encapsulation,
restricting and granting access to resources and inter-app
communication.

These packages are not portable across platforms and tied to the
respective app store. Additionally, they are designed for close
interaction with the provided platform, and there are usually
restrictions in place preventing the use of these platforms for general
software development.

\subsection{Summary}\label{summary-2}

It was shown that these approaches to software distribution have certain
disadvantages. Different abstraction levels have been introduced to
counter problems and provide more features.

However, using the shown approaches it is difficult to distribute and
deploy applications without distributing the source, cluttering the file
system and at the same time being portable and maintainable in data
centres.

\section{Existing Approaches}\label{existing-approaches}

Since the Docker software was first introduced engineers have developed
different approaches to containerize the building process and the
resulting software. A selection is listed here.

\subsection{Plain Dockerfile}\label{plain-dockerfile}

The first approach using the Docker platform is building the software
inside the container it is destined to run in. This can be accomplished
with Dockerfiles: A suitable base image or ancestor layer containing a
builder is sufficient.

\begin{lstlisting}[language=Dockerfile, caption=Dockerfile for in-container compilation, float=tbph, label=lst:dockerfile-with-gcc]
FROM ubuntu:latest
RUN apt-get update &&\
  apt-get install -y --no-install-recommends gcc &&\
  mkdir /source
WORKDIR /source
CMD ["/source/hello"]
ADD hello.c /source/
RUN gcc -o hello hello.c
\end{lstlisting}

An example for this approach can be found in listing
\ref{lst:dockerfile-with-gcc}. It is based on the latest Ubuntu base
image, and installs the GCC compiler. The \lstinline!WORKDIR! and
\lstinline!CMD! instructions provide defaults for later execution steps.
Afterwards, the source code file \lstinline!hello.c! is copied into the
image, and compiled in the next step. At this point, the image is
complete, and prints out the `Hello, World!' message on execution.

\subsubsection{Mathematical Formulation}\label{mathematical-formulation}

As was already shown, each command line in a Dockerfile adds a layer on
top of the existing layer stack. In this example, the command in line
eight, namely the execution of the compiler, can be considered a builder
according to the definition above. The input of this builder is the
image at the `time' of line seven. This layer is called \(i_7\). The
compilation step can then be expressed using the functions defined
above: \(i_8 = f_\xi(\mathrm{file}_{i_7})\).

Generally, each builder gets the directory created by the layer above,
transforms it according to its rules and returns a new directory.
Recursively applied results in
\(d_{n+1} = f_{\xi_{n+1}}\left(d_n\right)\), with \(d_0\) either a
directory with some content or an empty directory.

For a Dockerfile with three commands and a base directory \(d_0\) the
following equation holds:
\(\hat{d} = f_{\xi_3}\left(f_{\xi_2}\left(f_{\xi_1}\left(d_0\right)\right)\right)\).
The persisted image layers are given by the difference between two
intermediate directories, i.e.~by

\begin{align*}
  i_2 & = \mathrm{diff}\left(d_1, d_2\right) \\
    & = \mathrm{diff}\left(d_1, f_{\xi_2}\left(d_1\right)\right)\\
    & = \left.f_{\xi_2}\right|_{d_1}
\end{align*}

The stack produced by these layers is exactly the list of changes
introduced by the builder functions: \(I = (i_1, \dots, i_n)\). However,
the combined list of changes \(\hat{i} = i_1 \circ \dots \circ i_n\) is
not the same as the difference between the base directory and the file
system a container is able to see, i.e.~in general
\(\hat{i} \ne \mathrm{diff}\left(d_0, \mathrm{file}_I\right)\).

The image layer stack is generally bigger in terms of amount of data
than the change set between the base directory and the consolidated view
on the image layer stack. This is easy to see if the image stack layer
is considered to be the \emph{path} and the change set to be the
\emph{distance} to the final file system view.

\subsubsection{Discussion}\label{discussion}

The `Best Practices for Dockerfiles' by Docker Inc. recommend using as
few layers as possible, but not sacrificing long term maintainability
(see \autocite{dockerfileBestPractices}). However, there are upper
bounds to what can be achieved: Multiple consecutive \lstinline!RUN!
invocations can be coalesced into one by joining the commands into one
(using shell scripting). Similarly, consecutive \lstinline!ENV! commands
can be collapsed as well.

By coalescing multiple \lstinline!RUN! commands it is possible to save
space in the generated image, if one of the latter removes files
generated by one of the former. For example, it is quite common to
install software using a package manager such as Apt into a container,
but first populating the package cache prior to installation. If a
command that discards the cache afterwards is executed in the same
container, i.e.~in the same \lstinline!RUN! command, the package cache
files will leave no trace in the final image. An example for this is
shown in listing \ref{lst:coalesce-run-dockerfile}.

\begin{lstlisting}[language=Dockerfile, caption=Coalescing RUN commands, float=btph, label=lst:coalesce-run-dockerfile]
FROM ubuntu:trusty
# results in three additional layers,
# and the baggage of the apt-get cache files
RUN apt-get update
RUN apt-get install -y mtr
RUN apt-get clean &&\
  rm -rf /var/lib/apt/lists/* /tmp/* /var/tmp/*
# one layer, no baggage
RUN apt-get update &&\
  apt-get install -y mtr &&\
  apt-get clean &&\
  rm -rf /var/lib/apt/lists/* /tmp/* /var/tmp/*
\end{lstlisting}

In cases where different \lstinline!RUN! invocations need to be executed
using different user contexts or with different files copied from the
host it is not possible to coalesce them further. This limits the
optimizations that are possible and establishes a lower bound for the
image layer count. Additionally, the optimized Dockerfiles contain long
lines with multiple commands, optionally spread over multiple lines.

Another advantage of the build component of Docker that executes the
Dockerfiles is lost as well: Since the update, installation and cache
removal is done in one step, no caching between steps can help reduce
the network traffic incurred by the step. A Dockerfile using the first
part in listing \ref{lst:coalesce-run-dockerfile} reuses the package
files in later steps, while a change to just one subcommand in the
second part triggers a full re-execution of the whole step.

This behaviour incurs significant additional traffic when compiling
software with a package installed and later removed using this method.
For example, the Dockerfile for the official \lstinline!golang!
repository in version 1.6 in the \lstinline!alpine! variant installs
build dependencies, compiles the \lstinline!golang! sources, and then
removes the dependencies again in one step (see the Dockerfile by Gravi
\emph{et al.} \autocite{golangAlpineDockerfile}). Since the tagged
sources of \lstinline!golang! don't change, this step only has to be
executed once for the official Docker image build, but this approach is
not viable for frequently changing sources, i.e.~when software is
developed.

\subsection{Squash-And-Load}\label{squash-and-load}

During the construction of an image with a Dockerfile a stack of
internal layers between the base image and the layer that is at the top
of the stack when the build terminates may accumulate. Due to the
immutable stack architecture each layer is only able to \emph{add}
change sets, but can't remove any.

When building the image described in listing
\ref{lst:dockerfile-with-gcc}, the package manager has to install the
compiler, which is executed only once and then left in the image.
Deleting the compiler package in a later step is not sufficient since
this deletion only records this for new layers, and doesn't affect any
below.

The Docker software provides a direct way to get an archive containing
the files as they can be seen by a program running in a container
derived from an image. This view flattens the stack to a single change
set, any related additions and deletions cancel each other out.

For example, Uhrig recommends flattening the file system by creating a
container and exporting its view of the file system, and creating a new
image from this (see \autocite{dockerFlatten}). During this process any
meta data is removed, including the executed command and environment
variables, as well as the history of the image.

Jason Wilder offers a tool\footnote{\url{https://github.com/jwilder/docker-squash}}
that also squashes multiple layers together, removing any trace of
already deleted files. In difference to Uhrigs approach this tool
extracts the meta data before squashing the layers, and adds them to new
image.

\subsubsection{Mathematical
Formulation}\label{mathematical-formulation-1}

This approach uses the view a container has on the underlying file
system, and persists this in a new image. This can be formalized as
follows:

Let \(c_i\) be a container \(c_i\) based on image \(i\). The mapping
from file name \(f\) to file is then given by \(\mathtt{file}_i(f)\)
(see definition above). The new image layer \(l^\prime\) is then
generated such that \(l^\prime(f) = \mathtt{file}_i(f)\) if and only if
\(f \in \mathrm{dom}\left(\mathtt{file}_i\right)\).

\subsubsection{Discussion}\label{discussion-1}

This approach solves the accumulation of intermediate image layers. Only
the result of the last builder is taken and encapsulated as an image.
Applications with vastly different compile and runtime environments are
able to profit greatly from this approach, for example a Go tool needs a
tool suite weighing around \(750\) MiB for compilation, but the result
may very well be below the \(10\) MiB threshold.

An implementation detail is the loss of meta data specified in the base
image or the intermediate layers. Tooling can assist the repackaging
step, as can be seen by the multitude of tools already available.

Although only the files visible in the lastly used container are
extracted and packaged this approach is inefficient if a significant
part of these files remains constant over different versions of the
image, or is used in other images within the same host or environment.
For example, each Java based application ships its own virtual machine
with each update. It is computationally hard for tools to detect and
efficiently compress shared files in different images.

\section{Proposed Approach: Utility Containers plus Layer
Donning}\label{proposed-approach-utility-containers-plus-layer-donning}

\textbf{Utility Containers} employ the capabilities of a list of images
to transform a directory \emph{external} to the running container, for
example an image containing the GNU Compiler Collection (GCC)
transforming C source code into an executable binary. This concept
differs from the previously mentioned approaches by having the utility
in the container, but not the transformed code.

Sometimes, the result of a build process is a Docker image as well. To
produce an executable image \textbf{Layer Donning} makes use of the
layering capabilities Docker provides. A base image containing the
necessary runtime environment is used, and extended with a new layer on
top. The resulting image is a deliverable image containing both the
runtime environment and the executable.

Layer Donning differs from the building with Dockerfiles in that it only
uses one additional layer on top, while building with Dockerfiles adds
additional layers for each meta data change, for each executed command
and for each added file. It differs from the squash-and-load approach in
making use of the layering capabilities provided by Docker by attaching
a new layer containing the build result on top of an existing base
image.

\subsection{Mathematical Formulation}\label{mathematical-formulation-2}

The Layer Donning approach can be formalized using the mathematical
notation from above. In each step, the build tool uses one containerized
builder, mounts the source directory into the file system of the
container and uses the resulting directory for the next builder. Again,
\(d_0\) is considered to be the base directory.

Each step transforms the resulting directory according to the builder
\(f_{\xi_n}\). For all \(n\) greater than \(0\), the directory in
iteration \(n\) is given by the inductive definition
\(d_n = f_{\xi_{n+1}}\left(d_{n-1}\right)\).

Finally, an image can be created by wrapping any directory \(d^\prime\)
(possibly a subdirectory) into a base image \(i\). The new image will
consist of the image layers comprising \(i\) plus the layer created from
\(d^\prime\). In the traditional Docker image storage system this will
be \(i^\prime\) with \(\mathrm{parent}\left(i^\prime\right) = i\). In
the content addressable universe the hash of the new image layer is
appended to the list of hashes in \(i\) and stored as a new image
\(i^\prime\):
\(h_{c_{i^\prime}} = (h_{c_i1}, \dots, h_{c_in}, hash(i_x))\) with
\(file_{i_x} = d^\prime\).

\section{Implementation}\label{implementation}

To demonstrate the applicability of the proposed approach to real-world
software development a prototype called Involucro has been realized.
This section introduces the concepts and architecture of this software.

\subsection{Concepts}\label{concepts}

Involucro is configured with a Lua program. In this program, the
developer specifies \emph{steps} that have to be taken to build the
script. These steps are grouped into \emph{tasks} that can be called
from the command-line interface (but also from other tasks).

The available step types are:

\begin{description}
\tightlist
\item[run]
Runs a Docker container with the current directory made available.
\item[wrap]
Creates a new image with an optional base image and a specified
directory on top. The resulting image is always tagged into a
repository, the name is specified as part of this step.
\item[run task]
Executes another task defined in the same file. This can be used to
group some tasks together to form an overall build task.
\item[tag image]
The name assigned during wrapping is possibly not the only one an image
has. This step type allows assigning another repository name and tag to
an already existing image.
\item[hook]
This step allows the execution of a Lua function in-between steps. Most
importantly, this function is executed after other steps have already
been run. It is able to modify the build script using results from
previous steps.
\item[push]
The canonical way for Docker image distribution is via a registry. This
step type allows uploading an image into a registry, making it available
for download from other authorized hosts.
\end{description}

\subsection{Architecture}\label{architecture}

Involucro communicates directly with a Docker daemon. This means it is
not necessary to have a Docker client installed on the machine. To
increase portability across development machines and due to the
widespread adoption in the container community the programming language
Go\footnote{\url{https://golang.org/}} developed by Google was chosen to
implement the software.

The software works in two phases: The control file is read and executed
first. After all tasks have been defined the tasks requested by the user
are executed in order. The set of tasks can be changed during the second
phases using the hook step. This allows the definition of tasks whose
definition relies on the prior execution of other tasks. For example, a
compressed file has to be unpacked first before a task that resizes all
images can be defined.

Containers are executed in the same way as the native
\lstinline!docker run! command works: The container is created from an
image and started with a command. By default, Involucro mounts the
current working directory into the container at the path
\lstinline!/source! and sets the working directory for the container to
that path as well. Most containerized tools can be used without further
configuration with this model. Additional configuration options can be
set for the container and for the hosting environment in the control
file.

In theory it is possible to import a complete image composed out of
multiple image layers directly into the Docker service. Some versions of
Docker even allowed uploading only parts of the image: The import only
succeeded if omitted image layers were already present on the system,
`new' image layers were able to declare a parent-child relationship to
another image layer. When Docker implemented content addressable image
layers the possibility to inspect the layers comprising an image was
removed, making the creation of a multi-layered image archive based on
an existing image impossible at the time of writing.

The alternative chosen by Involucro is instantiating a container from
the new base image, copying the contents into the desired location and
committing it to a new image. This is the method the native Docker
builder uses as well which makes this method future-proof. It has the
disadvantage of copying the files comprising the new image layer once
into the container and then again into the new image layer when
committing but the cost of the extra copy is hardly noticeable for
normal sized image layers.

\subsection{Example}\label{example}

Listing \ref{lst:example-invfile} is an example for a control file
involving different kinds of steps. The build system described by this
file compiles a dynamic set of single-file C programs into executable
binaries and packs them together into a common image and tests it.

\begin{lstlisting}[style=mylua, caption=Example invfile.lua, float=bthp, label=lst:example-invfile]
inv.task('all')
  .runTask('prep')
  .runTask('gen:file_list')
  .runTask('gen:tasks')
  .runTask('compile')
  .runTask('package')
  .runTask('test')

inv.task('prep')
  .using('busybox:latest').run('mkdir', '-p', 'dist')

inv.task('gen:file_list').using('busybox:latest')
    .run('/bin/sh', '-c', 'ls -1 *.c |'
      .. 'xargs -I+ basename + .c > .files')

local compile_all = inv.task('compile')
inv.task('gen:tasks').hook(function ()
    for l in io.lines('.files') do
      inv.task('compile:' .. l)
        .using('frolvlad/alpine-gcc:latest')
          .run('gcc', l .. '.c', '-o', 'dist/' .. l)
      compile_all.runTask('compile:' .. l)
    end
  end)

inv.task('package').wrap('dist').inImage('alpine:latest')
    .at('/usr/local/bin').as('demo/toolset:v1')

inv.task('test').using('demo/toolset:v1')
    .withExpectation({stdout = "/usr/local/bin/a"})
    .run('/usr/local/bin/a')
\end{lstlisting}

The control file generates a list of files by invoking \lstinline!ls!
and stripping the suffix \lstinline!.c! to get the name of the
executable. This file list is written to the hidden file
\lstinline!.files! which is later read by the hook function executed as
part of the \lstinline!gen:tasks! task. This task generates the
compilation task for each of the source code files and registers it for
execution at the generic \lstinline!compile! task. The image is created
using the executables in the \lstinline!dist/! directory. Afterwards the
image is tested by invoking one of the programs and checking if the
output matches the given pattern. The program \lstinline!a! in this
example prints out the name it is called with, in this case
\lstinline!/usr/local/bin/a!. All sub-tasks are executable either by
their name or by invoking the main task \lstinline!all!.

\section{Evaluation}\label{evaluation}

In this section three test programs are introduced and implemented using
the described approaches. The approaches are evaluated against a set of
criteria and the results discussed.

\subsection{Test programs}\label{test-programs}

The evaluation was done for a set of programs, differing in size and
compilation and runtime complexity. They are introduced here.

\subsubsection{\texorpdfstring{`Hello, World!' with
Busybox}{Hello, World! with Busybox}}\label{hello-world-with-busybox}

The first and simplest example is to use a Busybox\footnote{Busybox is a
  combination of tiny versions of several common UNIX utilities. It is
  primarily used in embedded environments, where disk space is usually
  constrained. (\url{http://www.busybox.net/}).} image. It contains an
\lstinline!echo! tool, which is used to display the greeting `Hello,
World!' To display this message, a shell script in the image will be
used. The shell script is shown in listing \ref{lst:hw-shell-script}.

\begin{lstlisting}[language=bash, caption=Shell Script displaying 'Hello World', label=lst:hw-shell-script]
#!/bin/sh

echo "Hello, World!"
\end{lstlisting}

\subsubsection{Number Factorization in
C}\label{number-factorization-in-c}

The second example is a small tool calculating all prime factors of a
positive number greater than one. The source code written in C is shown
in listing \ref{lst:factorizer}.

The program reads one number from the standard input and divides it by
all dividers until the number is reduced to \(1\), at which point the
program can stop. It uses a very inefficient algorithm and could be
optimized easily, but is intentionally kept short and readable for this
context.

\begin{lstlisting}[language=C, caption=Factorizer, float=tbph, label=lst:factorizer]
#include <stdio.h>
int main() {
  int target = 0;
  if (scanf("%d", &target) < 1 || target <= 1) {
    fprintf(stderr, "Invalid number, should be > 1\n");
    return 1;
  }
  printf("%d:", target);
  int divider = 2;
  while(target > 1) {
    while (target % divider == 0) {
      printf(" %d", divider);
      target /= divider;
    }
    divider++;
  }
  printf("\n");
  return 0;
}
\end{lstlisting}

\subsubsection{Blog with Backend and
Frontend}\label{blog-with-backend-and-frontend}

As a third example a simple blog was used. It is implemented as a
two-tier web application with a backend server process and a frontend
web interface. With this example real-world implications of the
approaches can be explored.

\paragraph{The Backend Tier}\label{the-backend-tier}

Managing and controlling accesses to the storage system is the
responsibility of the backend tier, as well as transforming the native
data structures to a format suitable for transmission to the frontend.
In this case, a HTTP server written in the Go programming language is
used to query data from the database, and send it to the frontend.

Usually an access control component limits certain manipulations of the
data to authorized users only, but for simplicity this backend allows
all operations.

\paragraph{The Frontend Tier}\label{the-frontend-tier}

The frontend is implemented as a single-page application with the
Ember.js\footnote{\url{http://emberjs.com/}} framework. Described as `a
framework for creating ambitious web applications', it provides an
overall architecture for the development of large and complex
applications, as well as managing and rendering the data.

Following the advice by the developers of Ember.js the build tool
\lstinline!ember-cli! is used to handle assembling the final production
files as well as providing a development environment with incremental
builds and automatic reloads.

The code for both components is available in the supplement which is
referenced to in appendix \ref{supplement}.

\subsection{Criteria}\label{criteria}

In order to evaluate the proposed approach a set of evaluation criteria
was defined. These criteria are shown below:

\begin{description}
\item[Initial Compile: Elapsed Time / Traffic]
This criterion measures the time and bytes transmitted over the network
that are required for a full compilation of the example if no prior
results are available on the system of the user. For new developers this
is the most critical criterion as it is the amount they have to wait for
to get a completely functional tool. The time and network traffic is
measured without any Docker images or containers present on the system.

The available downspeed has an enormous influence on the time required
to build an image. It is difficult to accurately compare different
approaches by time if they rely on different remote sources to download
the needed tools. To mitigate this influence the actual network data is
measured and discussed as well.
\item[Re-Compile: Elapsed Time / Traffic]
After one clean build, the time and network usage is examined again for
the next build. A small number for this criterion means faster feedback
for developers and testers of the software. It is measured just after
the initial compile process has been executed, i.e.~it can use any
caches or files generated by the other step. Before the recompilation
one of the source code files is changed to force the compiler to
reexecute. The change does not affect the compilation speed of the file
(it doesn't change the complexity of the file) by changing the order of
commands or the values of string constants.
\item[Size of Deliverable]
In the end of the building phase there is a deliverable product that is
either downloadable by a client, or deployable into a data centre. It is
advantageous to have smaller deliverables when a user has to download it
via the Internet, but there is a similar situation in professional data
centres: Even though bandwidth and disk space are cheap it is still
desirable from a security point of view to have only the essential
amount of functionality in a container (as proposed by Saltzer and
Schroeder, \autocite{princLeastPrivil}).
\end{description}

\subsection{Realisation}\label{realisation}

In this section the introduced approaches will be applied to the three
exemplary projects.

\subsubsection{\texorpdfstring{A `Hello World'
Image}{A Hello World Image}}\label{a-hello-world-image}

The `Hello, World' image has to start the given script printing the
message, and exit afterwards.

\paragraph{With Dockerfile}\label{with-dockerfile}

A small Dockerfile is enough to create an image that exhibits the
desired behaviour. The one used in this evaluation is shown in listing
\ref{lst:eval-hw-dockerfile-dockerfile}.

\begin{lstlisting}[language=Dockerfile, caption='Hello World' Dockerfile, label=lst:eval-hw-dockerfile-dockerfile]
FROM busybox
ADD hello-world.sh /
CMD ["/bin/sh", "/hello-world.sh"]
\end{lstlisting}

\paragraph{Squash-and-Load}\label{squash-and-load-1}

An image displaying the `Hello, World!' message can be created with a
Squash-and-load approach by instantiating a container, squashing its
contents and loading it as a new image. At the same time, the default
command can be set to be the \lstinline!echo! command. Listing
\ref{lst:squash-and-load-hw} shows commands that can be used to create
such an image.

\begin{lstlisting}[language=bash, caption='Hello World' Squash-and-load, float=tbph, label=lst:squash-and-load-hw]
ID=$(docker create busybox)
docker cp helloworld.sh $ID:/
docker export $ID | docker import \
  -c "CMD /bin/sh /helloworld.sh" -
docker rm $ID
\end{lstlisting}

\paragraph{Layer Donning}\label{layer-donning}

Using the Involucro tool from above, the build step can be encoded using
the steps in listing \ref{lst:ld-hw}. The task can be invoked by running
\lstinline!involucro wrap!.

\begin{lstlisting}[style=mylua, caption='Hello World' using Involucro, float=htbp, label=lst:ld-hw]
inv.task('wrap')
  .wrap('.')
    .at('/')
    .withConfig({
      cmd = {"/bin/sh", "/hello-world.sh"}
    })
    .inImage('busybox:latest')
    .as('test/hello_world')
\end{lstlisting}

\subsubsection{Factorization Tool}\label{factorization-tool}

The factorization tool has to be compiled using a C compiler. When a
container is started, the executable should be invoked and connected to
the standard input and output.

\paragraph{Dockerfile}\label{dockerfile}

Using a Dockerfile to build the software, a C compiler has to be
installed into the container first. In line with the current
recommendations by Docker Inc. the \lstinline!alpine! image will be used
as base image. This image is very small (about \(3.2\) MiB), but still
provides a powerful and complete package manager. Two variants are
shown:

\begin{itemize}
\tightlist
\item
  The compiler is installed, executed and removed in one step after the
  source file is added to the image (listing
  \ref{lst:factorizer-dockerfile-a}). This has the advantage that no
  trace of the compiler remains in the final image, incurring no
  baggage.
\item
  The compiler is installed before the source file is added to the
  image. After the compilation step the compiler is removed, leaving
  only the executable behind (listing
  \ref{lst:factorizer-dockerfile-b}). This approach uses the caching
  provided by Dockerfiles.
\end{itemize}

\begin{lstlisting}[language=Dockerfile, caption=Factorization tool with Dockerfile (single step), float=htbp, label=lst:factorizer-dockerfile-a]
FROM alpine
CMD ["/factorizer"]
ADD factorizer.c /factorizer.c
RUN apk update &&\
  apk add gcc &&\
  gcc -O2 /factorizer.c -o /factorizer &&\
  apk del gcc &&\
  rm -rf /var/cache/apk/
\end{lstlisting}

\begin{lstlisting}[language=Dockerfile, caption=Factorization tool with Dockerfile (cached steps), float=htbp, label=lst:factorizer-dockerfile-b]
FROM alpine
CMD ["/factorizer"]
RUN apk update && apk add gcc
ADD factorizer.c /factorizer.c
RUN gcc -O2 /factorizer.c -o /factorizer &&\
  apk del gcc &&\
  rm -rf /var/cache/apk/
\end{lstlisting}

\paragraph{Squash-and-Load}\label{squash-and-load-2}

The squash-and-load approach can use one of the Dockerfiles from above.
There is no difference in the end result, but the second one can utilize
build caching.

\paragraph{Layer Donning}\label{layer-donning-1}

The factorization tool uses the utility image \lstinline!alpine-gcc! by
the user \lstinline!frolvlad! in the version \lstinline!latest!
available on Docker Hub\footnote{\url{https://hub.docker.com/r/frolvlad/alpine-gcc}}.
This image contains a C compiler with all necessary dependencies to
build software for an Alpine Linux installation. The control file
provides tasks to build the executable and to wrap it in an image for
distribution. It is shown in listing \ref{lst:factorization-invfile}.

\begin{lstlisting}[style=mylua, caption=Factorization tool using Involucro, float=htbp, label=lst:factorization-invfile]
inv.task('build')
  .using("frolvlad/alpine-gcc:latest")
    .run('/bin/sh', '-c', 'mkdir -p dist && '
      .. 'gcc -o dist/factorizer factorizer.c')

inv.task('package')
  .wrap('dist')
    .inImage('alpine:3.3')
    .at("/")
    .withConfig({cmd = {"/factorizer"}})
    .as("test/factorization")
\end{lstlisting}

\subsubsection{Blog}\label{blog}

The backend code of the blog engine implements a basic HTTP server for
the API calls. Using this API, posts and comments can be created,
updated, read, and deleted. It persists the corresponding records in the
database.

Prior to deployment, the frontend code is compiled into a reduced and
deliverable form which can be executed by a browser. The result of this
compilation step is a static HTML file and a set of associated
stylesheets and JavaScript files. During development, they are served by
the development tool \lstinline!ember-cli! allowing for automatic
reloading of the web application upon changes to the source code. In
production builds however these files are delivered by the blog engine.

\paragraph{Dockerfile}\label{dockerfile-1}

The image is based on the \lstinline!alpine! image as well. After all
(cacheable) meta data are set the packages from the \lstinline!alpine!
and from the \lstinline!npm! repository that are needed for compilation
of both parts are installed. For the backend part the Go compiler is
required, the frontend calls for the \lstinline!ember-cli! tool as well
as the \lstinline!bower! code manager.

Before the frontend code is copied into the container the files
controlling the installation of additional dependencies are added and
the dependencies installed. Due to this, dependencies are cached when
the source code changes, and only reinstalled upon changes to the
dependency specifications. Afterwards, the backend code is copied into
the container, dependencies are fetched and the code is built. The
resulting Dockerfile is shown in listing \ref{lst:blog-dockerfile}.

\begin{lstlisting}[language=Dockerfile, caption=Blog Engine Dockerfile, float=htbp, label=lst:blog-dockerfile]
FROM alpine
EXPOSE 8040
WORKDIR /blog/
ENV GOPATH=/go
RUN apk --no-cache add go nodejs git &&\
    mkdir -p /blog/frontend/ && \
    npm install -g npm &&\
  npm install -g ember-cli bower

ADD backend/ /go/src/github.com/thriqon/backend/
RUN go build -o /blog/blog github.com/thriqon/backend/

ADD frontend/package.json /frontend/
ADD frontend/npm-shrinkwrap.json /frontend/
ADD frontend/bower.json /frontend/
RUN cd /frontend/ &&\
  npm install &&\
  bower --allow-root install
ADD frontend/ /frontend/
RUN cd /frontend/ &&\
  ember build -prod --output-path /blog/frontend/
\end{lstlisting}

For the criterion `Time to Recompile' the code is changed to reflect
work by developers. This example is `changed' twice: once with a
modification in the frontend and once in the backend.

\paragraph{Squash-and-Load}\label{squash-and-load-3}

The squash-and-load method uses a similar Dockerfile to create the
initial image. There is an added command in the end that removes all
non-essential files from the image. This command is shown in listing
\ref{lst:blog-squash-and-load}. Afterwards, the process creates a
flattened version of the image which is the final result.

\begin{lstlisting}[language=Dockerfile, caption=Blog Engine Squash-and-Load, float=thbp, label=lst:blog-squash-and-load]
# [...]
RUN npm uninstall -g ember-cli bower &&\
  apk del go nodejs git &&\
  rm -r /frontend /go /root/.npm /tmp/npm* /root/.cache/ \
  /usr/lib/node_modules/ /tmp/async-disk-cache/
\end{lstlisting}

\paragraph{Layer Donning}\label{layer-donning-2}

The two components of the application are compiled separately in
different tasks. The backend compilation using the official
\lstinline!golang! repository is executed first. It puts the generated
executable inside the \lstinline!dist/! directory which will later be
packaged into an image. The frontend is compiled using a small utility
image containing a NodeJS environment with the Ember build tool
preinstalled. The compiled application is written to a subdirectory in
the \lstinline!dist/! directory from which the static files will be
served by the backend. The control file for Involucro is shown in
listing \ref{lst:blog-layer-donning}.

Each time the frontend is to be built the dependencies published in NPM
and Bower are checked for updates and completeness. In a real world
application these checks would be moved out of the main recompilation
step as they are rather time consuming and network reliant. In this
evaluation however this step was included to be able to directly compare
the approaches without adding additional optimizations.

However, using Involucro allows programmers to use certain optimizations
in the build process when they know that certain steps are not needed to
be executed (again). In the blog example, most of the time in the
recompilation case was spent by the package manager validating the
installation of all frontend modules for completeness. This check can be
omitted if the developer knows there were no changes.

\begin{lstlisting}[style=mylua, caption=Blog Engine Layer Donning, float=btph, label=lst:blog-layer-donning]
local package = "github.com/thriqon/blog"
inv.task('build:server').using('golang:1.6')
  .withConfig({
    env = {"CGO_ENABLED=0"},
    workingdir = "/go/src/" .. package,
  })
  .withHostConfig({binds = {
      "./backend:/go/src/" .. package, "./dist:/dist"
  }})
  .run('go', 'build', '-o', '/dist/blog', './.')

inv.task('build:frontend')
  .using('thriqon/alpine-ember-cli:latest')
    .withConfig({entrypoint = {"/bin/sh", "-c"}})
    .withHostConfig({binds = {
        './frontend:/source',
        './dist:/dist'
     }})
    .run('npm install && bower install --allow-root')
    .run('ember build -prod --output-path=/dist/frontend')

inv.task('build')
  .using('thriqon/alpine-ember-cli:latest')
    .withConfig({entrypoint = {"/bin/sh", "-c"}})
    .run('mkdir -p dist')
  .runTask('build:server')
  .runTask('build:frontend')

inv.task('package')
  .wrap('dist').at('/srv')
    .withConfig({cmd = {"/srv/blog"}}).as(VAR.TAG)
\end{lstlisting}

\subsection{Execution Environment}\label{execution-environment}

The measurements were taken using a custom test application executed on
a \lstinline!1gb! instance in the \lstinline!fra1! region on Digital
Ocean using the CoreOS \lstinline!1000.0.0! image. Additionally, the
Involucro program described in section \ref{implementation} was
downloaded and made available for the test application. After each
measurement all remaining images and containers were removed from the
system to isolate the measurements.

The network measurements are gathered by summing the traffic generated
during the run of the experiments on all network interfaces as counted
by the Linux kernel. Each experiment was run multiple times and the
average number for each data point was used. This procedure was repeated
over multiple days to rule out transient influences.

\subsection{Results}\label{results}

\begin{table}[tbph]
          \begin{tabular}{@{\extracolsep{4pt}}l l r r r r r@{}}\hline
                    \multicolumn{2}{l}{Name} & \multicolumn{2}{c}{Initial Compile} & \multicolumn{2}{c}{Recompile} & Size \\
                    \cline{3-4}\cline{5-6}\cline{7-7}
                            & & KiB & ms & KiB & ms & KiB\\\hline
                    \hline\multicolumn{7}{l}{Hello, World}\\
                     & Dockerfile          &           748&     2,558&             0&       339&         1,086\\
                     & Squash and Load     &           748&     2,794&             0&       393&         1,086\\
                     & Layer Donning       &           748&     2,352&             0&       188&         1,086\\
                    \hline\multicolumn{7}{l}{Factorizer}\\
                     & Dockerfile a        &       119,067&     7,507&       116,719&     4,928&         5,343\\
                     & Dockerfile b        &       119,094&    11,207&             6&     1,540&        92,608\\
                     & Squash and Load     &       119,268&    10,552&       116,807&     8,306&         4,692\\
                     & Layer Donning       &        48,929&    18,754&             3&       782&         4,692\\
                    \hline\multicolumn{7}{l}{Blog}\\
                     & Dockerfile JS       &       431,817&   475,658&            73&    71,769&       473,635\\
                     & Dockerfile Go       &       430,215&   498,301&        68,845&   287,367&       473,635\\
                     & Squash and Load     &       433,128&   512,215&        68,965&   329,781&        12,700\\
                     & Layer Donning       &       410,628&   302,897&         3,089&    82,960&         8,736\\
                  \end{tabular}
                  \caption{Measurements}
                  \label{tab:measurements}
                \end{table}

                \def\diaghello--world{%
                  \pgfplotsset{enlargelimits=0.15,width=8cm,xticklabels={Dockerfile\\(Size: 1086 KiB),Squash and Load\\(Size: 1086 KiB),Layer Donning\\(Size: 1086 KiB)},xtick=data,ymin=0,tick label style={/pgf/number format/fixed,align=center}}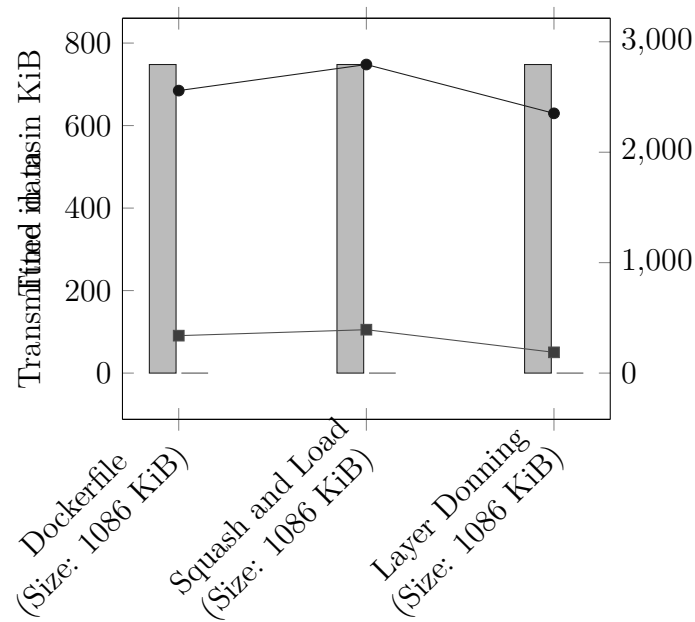
\begin{figure}[htbp]
                    \begin{tikzpicture}
                      \begin{axis}[ybar,axis y line*=left,ylabel={Transmitted data in KiB},x tick label style={rotate=45,anchor=east},scaled y ticks = false]
                        \addplot coordinates {(0, 748) (1, 748) (2, 748) };\label{plot:hello--world-ttc-kib}
                        \addplot coordinates {(0, 0) (1, 0) (2, 0) };\label{plot:hello--world-ttrc-kib}
                      \end{axis}
                      \begin{axis}[axis y line*=right,ylabel={Time in ms},xticklabels={},scaled y ticks = false]
                        \addplot coordinates {(0, 2558) (1, 2794) (2, 2352) };\label{plot:hello--world-ttc-ms}
                        \addplot coordinates {(0, 339) (1, 393) (2, 188) };\label{plot:hello--world-ttrc-ms}
                      \end{axis}
                    \end{tikzpicture}
                    \caption{Measurement results for 'Hello, World' example}
                    \label{fig:mes-res-hello--world}
                  \end{figure}
                }

                \def\diagfactorizer{%
                  \pgfplotsset{enlargelimits=0.15,width=8cm,xticklabels={Dockerfile a\\(Size: 5343 KiB),Dockerfile b\\(Size: 92608 KiB),Squash and Load\\(Size: 4692 KiB),Layer Donning\\(Size: 4692 KiB)},xtick=data,ymin=0,tick label style={/pgf/number format/fixed,align=center}}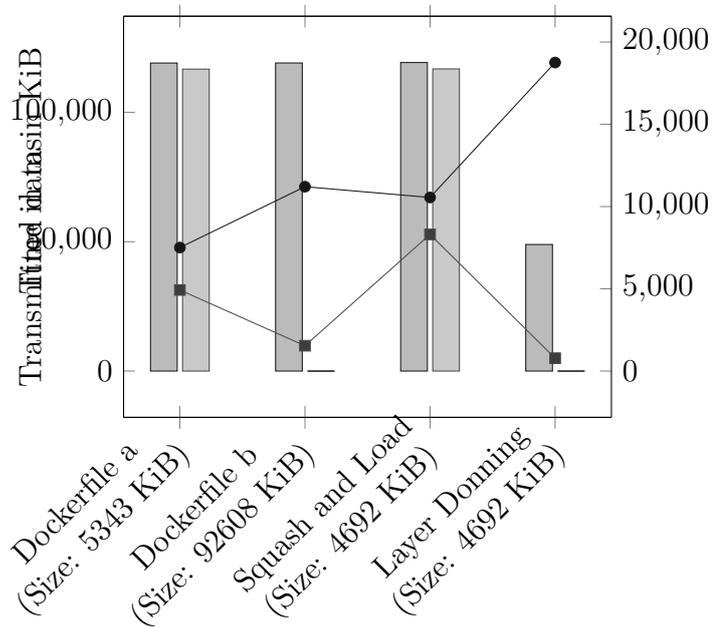
\begin{figure}[htbp]
                    \begin{tikzpicture}
                      \begin{axis}[ybar,axis y line*=left,ylabel={Transmitted data in KiB},x tick label style={rotate=45,anchor=east},scaled y ticks = false]
                        \addplot coordinates {(0, 119067) (1, 119094) (2, 119268) (3, 48929) };\label{plot:factorizer-ttc-kib}
                        \addplot coordinates {(0, 116719) (1, 6) (2, 116807) (3, 3) };\label{plot:factorizer-ttrc-kib}
                      \end{axis}
                      \begin{axis}[axis y line*=right,ylabel={Time in ms},xticklabels={},scaled y ticks = false]
                        \addplot coordinates {(0, 7507) (1, 11207) (2, 10552) (3, 18754) };\label{plot:factorizer-ttc-ms}
                        \addplot coordinates {(0, 4928) (1, 1540) (2, 8306) (3, 782) };\label{plot:factorizer-ttrc-ms}
                      \end{axis}
                    \end{tikzpicture}
                    \caption{Measurement results for 'Factorizer' example}
                    \label{fig:mes-res-factorizer}
                  \end{figure}
                }

                \def\diagblog{%
                  \pgfplotsset{enlargelimits=0.15,width=8cm,xticklabels={Dockerfile JS\\(Size: 473635 KiB),Dockerfile Go\\(Size: 473635 KiB),Squash and Load\\(Size: 12700 KiB),Layer Donning\\(Size: 8736 KiB)},xtick=data,ymin=0,tick label style={/pgf/number format/fixed,align=center}}\begin{figure}[tbp]
                    \begin{tikzpicture}
                      \begin{axis}[ybar,axis y line*=left,ylabel={Transmitted data in KiB},x tick label style={rotate=45,anchor=east},scaled y ticks = false]
                        \addplot coordinates {(0, 431817) (1, 430215) (2, 433128) (3, 410628) };\label{plot:blog-ttc-kib}
                        \addplot coordinates {(0, 73) (1, 68845) (2,  68965) (3, 3089) };\label{plot:blog-ttrc-kib}
                      \end{axis}
                      \begin{axis}[axis y line*=right,ylabel={Time in ms},xticklabels={},scaled y ticks = false]
                        \addplot coordinates {(0, 475658) (1, 498301) (2, 512215) (3, 302897) };\label{plot:blog-ttc-ms}
                        \addplot coordinates {(0, 71769) (1, 287367) (2, 329781) (3, 82960) };\label{plot:blog-ttrc-ms}
                      \end{axis}
                    \end{tikzpicture}
                    \caption{Measurement results for 'Blog' example}
                    \label{fig:mes-res-blog}
                  \end{figure}
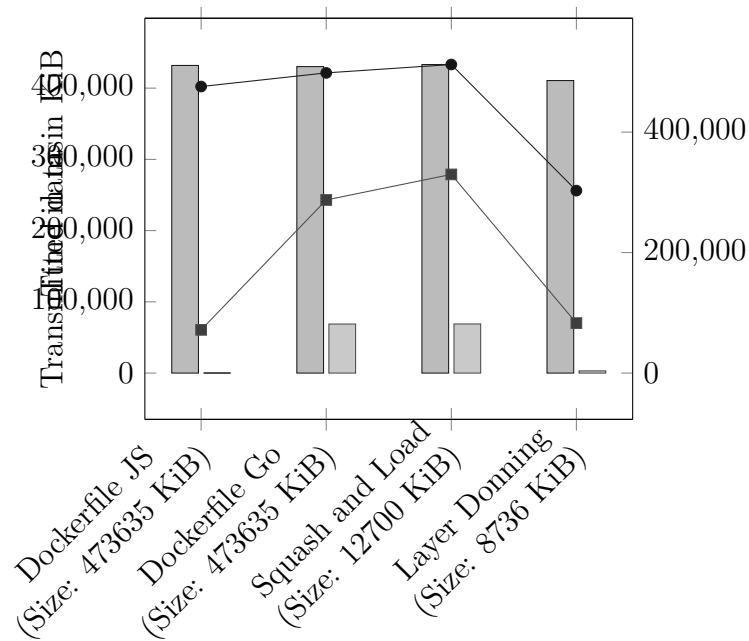
                }

The results of the evaluation program is shown in table
\ref{tab:measurements}. The numbers are visualized graphically in a
diagram for each evaluation example and discussed below.

\subsubsection{\texorpdfstring{`Hello, World'
Example}{Hello, World Example}}\label{hello-world-example}

In the first example, a Docker image that shows the `Hello, World'
message is created. All three approaches use similar commands to copy
the script into the image and set the command for execution. The results
are shown in diagram \ref{fig:mes-res-hello--world}.

On the first run, it is necessary for all three approaches to load the
base image from Docker Hub. This accounts for the
\ref{plot:hello--world-ttc-kib} \(747\) KiB of data used by all
approaches and for most of the \ref{plot:hello--world-ttc-ms} time
elapsed.

\diaghello--world

Afterwards, as a way to simulate a change to the programs code the
script is changed to display a different message (`Hello, Moon'). The
\ref{plot:hello--world-ttrc-ms} time and
\ref{plot:hello--world-ttrc-kib} data needed to recreate the image is
slightly smaller for the layer donning approach. This may be attributed
to the single-step encapsulation implementation of Involucro instead of
the two-step approach used by the native Docker builder. This difference
however is unlikely to surface in real-world applications.

The deliverable image has the same size in all three approaches
(\(1,086\) KiB).

\subsubsection{Factorizer Example}\label{factorizer-example}

The factorizer example uses two different exemplary Dockerfiles to build
the final image. The first version (\lstinline!a!) shown in listing
\ref{lst:factorizer-dockerfile-a} executes all installation and
compilation tasks in one container, while the second version
(\lstinline!b!) in listing \ref{lst:factorizer-dockerfile-b} separates
the installation and compilation steps to enable the use of an
intermediate cache. Both versions have to download a set of dependencies
needed to compile the source code into an executable. The amount of
\ref{plot:factorizer-ttc-kib} transmitted data is almost equal in both
instances, but the \ref{plot:factorizer-ttc-ms} time needed is higher
for the second version. This is due to the intermediate commits and
container creations.

\diagfactorizer

When the tool is recompiled after a change to the source code the second
version can use the cache to resume the image creation just after the
file has been added. Only the actual compilation is left to do. This
leads to a negligible amount of \ref{plot:factorizer-ttrc-kib}
transmitted data and a significantly lower \ref{plot:factorizer-ttrc-ms}
execution time. The cache that lowers the recompilation time is paid for
by an increased deliverable size. The deliverable produced by the
squash-and-load is a few hundred KiB smaller than the result produced by
the first Dockerfile version. This can be explained by the different
number of layers in the final image.

The layer donning approach uses less \ref{plot:factorizer-ttc-kib}
transmitted data but takes \ref{plot:factorizer-ttc-ms} longer in the
first run. This can be attributed to different download speeds from
different hosts. Measurements indicate a higher bandwidth when
downloading from the Alpine repository in comparison with Docker Hub
(around \(80\) MB/s in contrast with \(808\) KB/s).

The two Dockerfile versions and the squash-and-load approach each excel
in one category. The first version and the squash-and-load approach
produce a small deliverable at the cost of runtime and transmitted data
in a recompilation while the second version manages to compile the code
again with minimal repeated downloads. The layer donning approach
however is able to produce a deliverable of the same size as the
squash-and-load approach. Additionally, it takes less time than the best
Dockerfile-based approach.

\subsubsection{Blog Example}\label{blog-example}

The blog example is made up of two independent build targets. They are
compiled in sequence in all versions. In this example however the impact
of the one dimensional Docker build cache can be seen.

The initial build uses similar amounts of \ref{plot:blog-ttc-kib} data
and \ref{plot:blog-ttc-ms} time for all approaches. The increased time
needed for the squash-and-load approach might be attributable to the
additional cleaning steps which take non-trivial time.

The two experiments with the Dockerfile show the difference between
executions that occurs with the same Dockerfile (listing
\ref{lst:blog-dockerfile}) but with a different file change: In the
first experiment the code of the frontend is changed by the test
environment and the application is recompiled. The second experiment
changes the code of the backend and invokes the recompilation.

\diagblog

In the first case the Docker build cache enables the process to skip
downloading and installing the dependencies of the frontend again
because nothing was changed in the Dockerfile and all \lstinline!ADD!ed
files up to this point. Cached images are used to resume building at the
point of the first change in files.

The second example is able to use the Docker build cache as well but
only up to the point of the addition of the backend server code. With
the new file revision in place the previously used cache is invalidated
and all dependencies for the frontend have to be fetched again. This
cache invalidation explains the increased \ref{plot:blog-ttrc-kib} data
and \ref{plot:blog-ttrc-ms} time for the recompilation step in the
second example.

Similar to the factorizer example the squash-and-load approach fetches
the dependencies anew. Using more time it manages to decrease the size
of the deliverable to around \(2.5\%\) of the size produced by the
Dockerfiles. This additional time is spent removing the build
dependencies from the image and squashing it into one layer. Similar to
the factorization example, the deliverable produced by the
squash-and-load approach only contains one layer. This means that any
update to the image contents necessitates a full refetch of the whole
layer.

The squash-and-load approach is only tested with a change in the backend
code. A change in the frontend code will lead to a similar performance
characteristic as the first Dockerfile experiment.

The layer donning approach uses specialized images to compile the two
components. It utilizes the official Golang image for the backend and a
specialized \lstinline!ember-cli! image for the frontend. This allows
for a \ref{plot:blog-ttc-ms} speed improvement over the other
approaches, possibly attributable to the needlessness of calculating the
tree of dependencies when installing the tool directly from the
repository. The \ref{plot:blog-ttc-kib} transmitted data however is
similar.

In the recompilation case only very little \ref{plot:blog-ttrc-kib} data
is transmitted. This could be reduced further by omitting the
installation step for the frontend dependencies when developers exploit
their knowledge of changes in the dependency control file. By omitting
this step the \ref{plot:blog-ttrc-ms} time required for a recompilation
can also be further reduced.

The squash-and-load approach as well as the layer donning approach
produce images in the \(1\) to \(20\) MiB range. These images do not
have NodeJS installed because it is not needed to serve the blog. On the
other hand, the image build with a Dockerfile uses almost \(500\) MiB.
This includes a full installation of the Go compiler as well as a full
installation of NodeJS and the Ember compilation utility.

\section{Discussion}\label{discussion-2}

The results demonstrate that the layer donning approach uses less or
similar time and transmitted network data to produce a smaller or
similar sized deliverable image. This holds true for the recompilation
case as well. From a performance perspective the proposed approach can
accordingly be considered advantageous.

The control file for Involucro is an executable program in a
Turing-complete language offering a rich API to developers. This allows
full programmatic control over the build system, but this flexibility is
paid for by increased complexity and consequently a steeper learning
curve than a simple and easy to learn Dockerfile. However, the primary
design criterion of building software allows the assumption that the
majority users will have prior programming experience and welcome the
additional possibilities.

\subsection{Idiomaticity}\label{idiomaticity}

A container is started around a single process and its child processes.
As soon as that process dies the whole container is terminated. It is
possible to instantiate different applications with their processes
managed by a common supervisor process in a single container, but this
makes process management harder as there would be multiple process
hierarchies instead of one. This principle is expressed in the idiom
`Run only one process per container' \autocite{dockerfileBestPractices}.

The prevalent approaches to image construction promote using multiple
different programs inside containers instantiated from the same image.
This defeats the single-purpose design an image is supposed to have.
Having the functionality of a full package management system or a
compiler suite in an application image is usually neither necessary nor
beneficial from a security point of view. Single-purpose images tend to
be of lower size which additionally speeds up transmission steps and
reduces network congestion, making the transition to a containerized
software delivery system easier for developers.

The proposed approach uses specialized utility containers that do one
job only, and uses layer donning to create a new image with exactly one
purpose. This approach directly fulfils the requirements placed onto
idiomatic image creation.

\subsection{Reproducibility}\label{reproducibility}

Dockerfiles are often advertised in favour of other methods to create a
Docker image because it is supposed to make it possible to retake the
steps taken to build the image. Reproducible containers based on
reproducible images are the foundations for reproducible computations.
Scientific experiments conventionally attempt to make their results
reproducible by describing the exact state of all components to enable
other researches to check the results.

In the context software building and delivery, reproducibility means
that the exact same software has to be made available. Usually, the same
(developer managed) version number is used, but is preferable to make
sure that binaries in use are bitwise equivalent.

It is easy to see that only if the `ingredients' for a Docker image are
reproducible the whole image can be considered reproducible. However, if
any of the software components in an environment is installed via a
`normal' package manager such as Apt, it is possible that different
users get different package revisions when fetching the packages at
different times. This insecurity regarding the exact bitwise version is
forwarded to an image construction process, because packages installed
from a package manager are usually specified by name only, or are
downloaded from an uncontrolled remote host.

Dockerfiles provide no safety here: Without additional constraints
imposed by maintainers of the files they only document the
\emph{commands} used to build the image, not the actual `ingredients',
i.e.~the \emph{files} that make up the image. The official images
provided by Docker provide this safety, as the maintainers only accept
Dockerfiles that use checksummed files. This restriction does not apply
to general Dockerfiles, however.

The proposed layer-donning approach provides this safety by default: If
all used utility images are referenced by an immutable tag, for example
a checksum for a content addressed image, the resulting image is
reproducible from the build recipe in combination with the source files.
This can be seen from above: A sequence of reproducible builders
\(\left(f_\xi\right)_n\) forms a combined builder \(\hat{f_\xi}\) by
chained application of the intermediate results, such that
\(\hat{f_\xi} = (f_\xi)_1 \circ (f_\xi)_2 \circ \dots \circ (f_\xi)_n\),
which itself is reproducible.

Source code of software is usually kept in a version control system such
as Git\footnote{\url{http://git-scm.org}}. It is possible to get an
exact copy of a directory at a given time from a version control system.
In combination with reproducible builders the whole build workflow is
accordingly reproducible.

A containerized workflow using the proposed approach can use utility
containers instantiated from images uniquely identified by their
cryptographic hash. This assurance of bitwise equal compiler images is
not present when using a compiler executable installed from a
traditional package manager on a developer workstation.

\subsection{Containers and Package
Managers}\label{containers-and-package-managers}

Software delivery with containers brings up the question of the role of
package managers in a containerized environment. Traditionally, it has
been the task of package managers to make software available to users,
to supply them with updates and to manage the lifecycles of associated
files. These tasks are handled by container management tools as well.

An image encapsulates an application with all its dependencies, while a
package manager typically packages parts that compose the final
application. Common libraries are managed by the package manager and
made available to other packages. However, a library package on its own
can't constitute an executable container. Images are typically only
created for executable applications.

Widespread package managers have transparent mechanisms to ensure the
integrity of the packages with respect to the original sources, for
example with cryptographic signatures of the generated package files
(see Cappos \emph{et al.} \autocite{packageSecurity}). Some container
formats already include a safe approach to sign and verify the correct
and untampered transmission from a trusted builder, but it is vital for
the trust in container platforms to make it possible to reproduce the
image building process as well.

A study by Gummaraju \emph{et al.} discovered that around 30\% of the
official repositories in Docker Hub contain images that are vulnerable
to `a variety of security attacks' \autocite{hubvulns}. Their
recommendation is to frequently rebuild images with the latest software
revisions to alleviate the risk of running vulnerable software. On the
other hand, package managers in widespread use usually have some sort of
update mechanism to distribute patches to vulnerable software in a
reasonable amount of time.

\section{Cluster Deployment}\label{cluster-deployment}

After packaging software into a container the software can be made
available to the user for download, for example in a publicly available
registry. Software that powers network services however is often not
published for download but deployed into a data centre operated by the
developing company. Containers are particularly suited as a unit of
deployment because of their lightweight and standalone nature.

Google has been running their software in containers for over a decade
(see \autocite{burns}) using different container management systems.
These systems enable them to manage containers in the scale of `hundreds
of thousands jobs' \autocite{verma2015large} across their data centres.

Recently, Google published Kubernetes\footnote{\url{http://kubernetes.io/}},
a new and open source iteration of their container management software.
It handles scheduling containers onto nodes in a cluster. A set of
containers can form a \emph{pod}, which is always scheduled jointly.
Multiple instances of the same pod specification are managed by a
\emph{replication controller} and can provide a \emph{service}.
Stateless services control the scheduling of pods onto nodes according
to the specifications of replication controllers and services.

\chapter{Application: Mulled}\label{application-mulled}

Traditional approaches towards software distribution have certain flaws,
as was shown above. To demonstrate the practicability of using
containers as a means to deliver software a system for automated
packaging was developed and evaluated. This section introduces the
concepts and design decisions as well as the evaluation results by
members of the container community.

Mulled was designed to be a system for largely unattended building and
uploading of Docker images containing fully functional installations of
open source software. The only interaction required is the specification
of a Mulled image in a configuration file. After the build has
terminated, the finished image is ready for download by the user.

To combat the problem of non-standard build systems and complex
dependency management various package managers have been developed that
encode the build steps for one package in a standardized format similar
to a shell script. Mulled contains adapters that allow utilizing these
repositories of build scripts. After the packages have been installed in
a location, a Docker image is generated with the result and uploaded to
a public repository.

\section{Architecture}\label{architecture-1}

Mulled is controlled by a repository on GitHub\footnote{\url{https://github.com}},
containing the source code of Mulled and the package specification file,
formatted as a Tab-Separated-Values (TSV) formatted file. An example is
shown in listing \ref{lst:example-packages-tsv}. Each line in this file
specifies one package with four fields:

\begin{description}
\tightlist
\item[Packager]
This field sets the package manager that is used to install the package.
\item[Package]
The name of the package in the package manager and in the Mulled
repository.
\item[Revision]
Mulled allows multiple revisions of images to coexist at the same time.
With this field, it is possible to enumerate multiple revisions by
repeating the package line with different revision values. Some
packagers are able to explicitly install specific versions and interpret
this revision field directly.
\item[Test]
Unfortunately, packaging is not trivial, since all dependencies have to
be contained inside the package. However, before publishing it is
obviously desirable to verify the correctness of the package. This field
provides a shell script that is executed in the context of the new
image, and is able to execute any tests necessary to verify it.
\end{description}

\begin{lstlisting}[caption=Example `packages.tsv`, label=lst:example-packages-tsv]
# Packager Package Revision Test
conda      tmux    2.1--1   tmux -V
alpine     go      1        go version | grep 'go version'
\end{lstlisting}

\begin{figure}[htbp]
\centering
\includegraphics{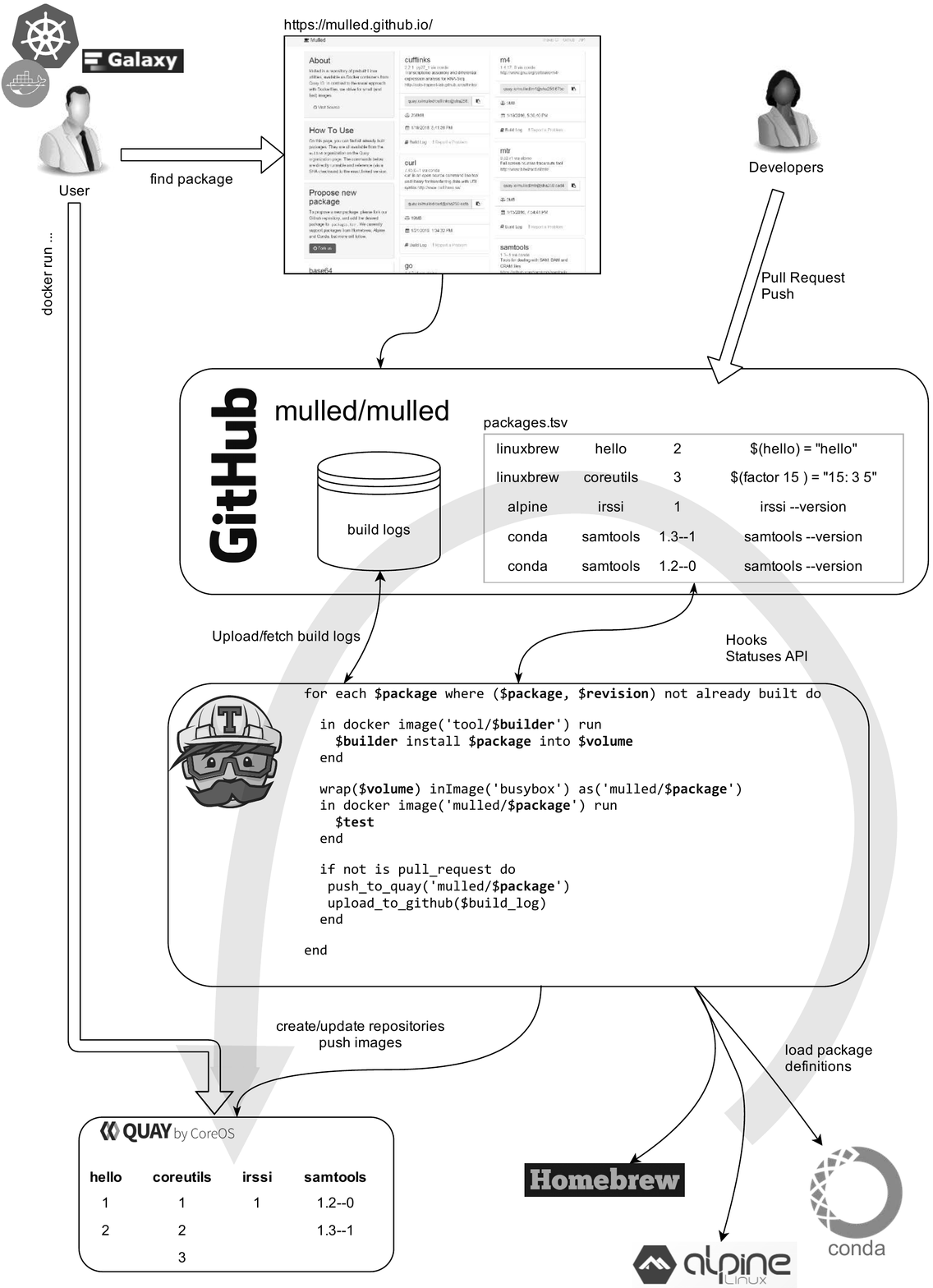}
\caption{\label{fig:mulledflow}Flow of information access in Mulled}
\end{figure}

When a commit is pushed into the GitHub repository, a build job is
created automatically on Travis CI\footnote{\url{https://travis-ci.org}}.
This job runs the build script and optionally pushes the result image
into a public repository on Quay.io\footnote{\url{https://quay.io}}.
This push only happens when the package specification has been accepted
into the repository, i.e.~when the commit appears on \lstinline!master!.
Figure \ref{fig:mulledflow} shows the flow of information in Mulled.

\subsection{Determination of build
targets}\label{determination-of-build-targets}

Any time the table changes on GitHub, the build script compares the
revision information stored in the data directory of the Mulled instance
(the \emph{actual} state) with the information stored in the table (the
\emph{desired} state). Additions in the desired state are resolved by
building the given package. These states are treated as sets of package
name and revision pairs. If the sets of package/revision pairs in the
actual state is \(A\) and \(D\) in the desired state, the packages that
need to be rebuilt are: \(R = D \setminus A\).

Another possible model is to evaluate the differences between two states
of the table by looking at the log of the version control system. The
advantage is that the build script does not rely on external systems and
is faster due to less network communication. However, if a build fails
for any reason, this may result in packages that are not built and
published, but also not easily rebuilt, since there is no build for them
that can be restarted.

\subsection{Choice of Technologies}\label{choice-of-technologies}

There are several publicly available systems that allow execution of a
build job when the code in a GitHub repository changes. These systems
are commonly used for Continuous Integration, but can also be used for
this use case: Continuous Integration services provide a neutral,
unbiased service that can be used to validate a new revision (of code,
for example) against a set of rules, for example a test suite (see Meyer
\autocite{meyerci}). In Mulled, the `change' to be tested is the
inclusion of a new package, and the `rule validation' is the successful
run of the build and test scripts.

Travis CI was chosen for this part of the system for the following
reasons:

\begin{itemize}
\tightlist
\item
  Available for open source projects for free: Projects that are hosted
  in public GitHub repositories are eligible for cost-free builds on the
  Travis CI infrastructure (see Travis CI plans \autocite{travisPlan}).
\item
  Travis CI supports a builder mode allowing for nested virtualization.
  This means that it is possible to have a fully functional Docker
  daemon inside a build job. Other continuous integration systems
  execute the build job itself in a Docker container, and do not allow
  spawning additional containers, which, however, is required to use the
  Involucro software.
\item
  Close integration with GitHub: Travis CI can be configured to build
  every proposed change as soon as it is submitted, and the result of
  this build is displayed in the GitHub user interface. Due to this
  automatic testing feature a repository administrator can safely accept
  new packages.
\end{itemize}

The default for Docker image hosting is directly encoded into the Docker
source code to be the Docker Hub. This default was not suitable for
Mulled because the offered API is lacking the important feature of
managing repositories programmatically. Quay.io on the other hand allows
full repository administration with authenticated network calls, and
(similar to Travis CI) offers a free tier for images that are publicly
available.

\section{Discussion}\label{discussion-3}

In theory, package managers provide their packages with proper
dependency meta data. It should be possible to install a package and
have all required code installed afterwards to use the software in a
reasonable way. However, it remains unspecified what pre-existing
software the package managers expect to exist in the target. For
example, during the development of Mulled it was discovered that a few
packages in the \lstinline!bioconda!\footnote{\url{https://github.com/bioconda/bioconda-recipes}}
repository depend on the \lstinline!zlib!\footnote{\url{http://www.zlib.net/}}
library to be present but did not depend on it in a `formal' way. These
bugs were not discovered earlier because the \lstinline!zlib! library is
available on most developer machines. In addition to providing a
repository of containerized software Mulled can therefore be used to
validate dependency specifications.

Each Docker image has to have a version string. In Mulled, this has to
be derived from the revision identifier specified in the control file to
enable efficient correlation of already built images and package
specifications. Unfortunately, only some package managers allow
installing specific versions of packages, the rest offers the most
recent version only. This makes reproducing previously built packages
difficult as the exact sources used for the build are depending on the
time the package is built. A builder that supports exact version
matching like the \lstinline!bioconda! builder based on the Anaconda
package manager is helpful here.

\chapter{Related Work}\label{related-work}

In this chapter other work for idiomatic and reproducible container
builds are reviewed.

\section{Packer}\label{packer}

HashiCorp develops Packer which aims to create `identical machine images
for multiple platforms from a single configuration'
\autocite{packer-intro}. It is able to create images in a wide range of
formats, including Docker.

The tool takes a configuration file and applies the commands on a
container instantiated from a base image. After the successful execution
the new container is saved as an image and optionally tagged into a
repository.

This type of packaging flow is an instance of the squash-and-load
approach, but without building the layers in between. From the
perspective of this thesis, it provides little benefit in terms of
efficiency (there is no caching between steps).

HashiCorp recommends provisioning a Docker image the same way a
conventional virtual machine is provisioned, for example using a
configuration management system such as Ansible or Chef. Treating a
container as a virtual machine is possible but not idiomatic. A
lightweight container should only contain one process and fulfil a
single purpose (see Melia \emph{et al.} \autocite{cisco-whitepaper}).

\section{Holy Build Box}\label{holy-build-box}

The Holy Build Box is a system for `building cross-distribution Linux
binaries' \autocite{holyBuildBoxReadme}. Due to inconsistencies in the
location and exact version of libraries and configuration files it is
traditionally problematic to run binaries compiled under another Linux
distribution. The Holy Build Box attempts to solve this by statically
linking most libraries except those that are expected to be present on
all target machines. This strategy enables finding the `sweet spot'
between full static and full dynamic linking.

Unfortunately, this approach only allows distribution of binaries that
can take the form of Linux binaries. Other types of deliverables are not
covered by this tool.

\section{AppImageKit}\label{appimagekit}

AppImageKit\footnote{\url{http://appimage.org}} was developed to solve
the problem of packaging additional resources for an application, for
instance libraries and images. It allows the maintainer of a software
product to create one distributable file that is executable across a
wide range of Linux distributions.

Internally, the kit uses a small setup utility which mounts an ISO9660
file system image (conventionally used for CD-ROMs) containing the
executable and its resources. Updates can be applied incrementally using
an official tool by the developers.

AppImageKit is designed to enable distribution of applications onto the
computers of end users. Currently, it does not employ additional sandbox
technology that would isolate the program from the host system which
enables an end user to work with the program exactly the same,
i.e.~their files are accessible in the usual locations. On the other
hand, this limits the usefulness in environments where applications
should be separated cleanly from each other, for example to fulfil
compliance policies.

\section{Global Alliance for Genomics and Health Data Working
Group}\label{global-alliance-for-genomics-and-health-data-working-group}

The Global Alliance for Genomics and Health (GA4GH) Data Working Group
(DWG)\footnote{\url{http://ga4gh.org/\#/}} is a part of the Global
Alliance\footnote{\url{https://genomicsandhealth.org/}} for Genomics \&
Health. It has a team tasked with `coordinating efforts around the
development of languages for describing repeatable genomic workflows'
\autocite{ga4gh-cwf-site}. They are in the process of specifying the
Common Workflow Language (CWL)\footnote{\url{http://www.commonwl.org/}}
based on YAML which can be used to specify a set of parameters to
control execution of tools. The writer of a workflow is able to flexibly
specify input and output files for the processes. The execution engine
is tasked with moving the files to the correct location for each step in
the workflow according to the specification.

As an option, the tools in the workflow can be encapsulated in a Docker
image. This enables complex workflow including many diverse images. An
exemplary workflow using Docker images from the user guide for CWL
\autocite{cwl-user-guide} is shown in listing \ref{lst:cwl-workflow}.
This workflow uses the official Java image to compile the file with the
id `src' and marks the files with the suffix \lstinline!.class! i.e.~the
generated class files as the output files.

\begin{lstlisting}[language=yaml, caption=Workflow in CWL for Java compilation, float=tbhp, label=lst:cwl-workflow]
cwlVersion: cwl:draft-3
class: CommandLineTool
baseCommand: javac
hints:
  - class: DockerRequirement
    dockerPull: java:7
    baseCommand: javac
arguments:
  - prefix: "-d"
    valueFrom: $(runtime.outdir)
inputs:
  - id: src
    type: File
    inputBinding:
      position: 1
outputs:
  - id: classfile
    type: File
    outputBinding:
      glob: "*.class"
\end{lstlisting}

There is a similar language called the Workflow Description Language
(WDL)\footnote{\url{https://github.com/broadinstitute/wdl}} in a custom
format providing comparable to define workflows. With both languages
complex workflows involving multiple different tools can be described.
They are particularly suited for multi-tool data processing for which no
special tool exists.

In software development however it is common to have a specialized build
tool for each language or environment that is able to exploit deep
knowledge about the structure of the code, for example to speed up the
compilation and to use simple configuration. The build tool
\lstinline!ember-cli! used in the blog example is such a specialized
utility.

When comparing the workflows approach by the GA4GH DWG with the approach
discussed in this thesis it is apparent that the workflows correspond to
utility containers. Both approaches provide reproducibility by uniquely
identifying code and environment for each step in the workflow as well
as a self-documenting task description.

Creating new images from source code is not in the scope of the workflow
approach and there is no indication that is planned to be included in
the future.

\chapter{Conclusion}\label{conclusion}

The main purpose of this study was to find a method to create Docker
images satisfying the idiomaticity and reproducibility criteria, and to
compare it with the methods prevalent in the container community today.
A new approach was proposed that solves the problems discovered in the
other models. The performance was validated using a new software that
applies this approach and a set of examples with varying complexity, and
with the automated Docker image creator Mulled.

The results of the evaluation and the positive results of Mulled
validate the approach with respect to performance and applicability:
Images are created faster, with less network traffic incurred and are
smaller than their counterparts created with legacy approaches. All
three factors are important when considering the introduction of
containerization into the software building and deployment workflow.
Great differences in the size and functionality between compile and
runtime environments call for a differentiation in their containerized
counterparts, and this separation is only provided by the Layer Donning
approach.

Mulled will be introduced to the scientific as an Application Note in
the near future. A relevant paper is in preparation. End users can use
Mulled today to create images for a great number of software packages
and with the help of a graphical tool like Kitematik\footnote{\url{https://kitematic.com/}}
these images can be executed on their computers without having to resort
to the command line. Organisations like iPlant\footnote{\url{http://www.iplantcollaborative.org/}}
can utilitze a Mulled instance to automatically derive Docker images
from existing software repositories without having to write down
installation steps for each package by hand.

The non-linear nature of the workflows described by Involucro control
files make it possible to execute steps in parallel if it is deemed
appropriate in an application. Further work in Involucro is needed to
support this requirement. It is currently out of the scope of Involucro
to support any kind of dependency-tracking which would support
automatically determining steps to be executed. It is however one of the
most frequently asked for features and could increase the adoption of
Involucro as full development system.

A close integration into the Bioconda repository is currently in
progress which will allow having a companion Docker image for each
package as soon as it is admitted into the repository. In the near
future, it could be up to the user or researcher whether to use a
package or an equivalent container image to support their work.

\appendix

\chapter{Supplement}\label{supplement}

Supplemental material can be found on GitHub in the repository
\url{https://github.com/thriqon/thesis-supplement}. The version of the
code used in this thesis is committed with id
\lstinline!5398f1d34101ebb46544744be2f540f9e349a905!, also available
under the tag \lstinline!release! signed by the PGP key
\lstinline!A6EBEF162E480D7E!.

The following materials are available there:

\begin{itemize}
\item
  The Involucro source code
\item
  The evaluation environment

  \begin{itemize}
  \tightlist
  \item
    Testing Engine
  \item
    Implementations for all three examples
  \end{itemize}
\item
  The evaluation results
\item
  The Mulled source code
\end{itemize}

\backmatter


\chapter{References}\label{references}
{\raggedright
\printbibliography[heading=none,category=cited]
}

\end{document}